\documentclass[twocolumn,showpacs,nofootinbib,superscriptaddress]{revtex4-1}
\usepackage{amsmath}
\usepackage{amssymb}
\usepackage{graphicx}

\DeclareMathOperator{\re}{Re}
\DeclareMathOperator{\im}{Im}

\begin{document}

\title{Dirac fermions in strong electric field and quantum transport in
graphene}

\affiliation{Institute of Physics, University of S\~{a}o Paulo, CP 66318,
CEP 05315-970 S\~{a}o Paulo, SP, Brazil}

\affiliation{Department of General and Experimental Physics, Herzen State
Pedagogical University of Russia, Moyka emb. 48, 191186 St. Petersburg,
Russia}

\author{S. P. Gavrilov}
\email{gavrilovsergeyp@yahoo.com}

\affiliation{Institute of Physics, University of S\~{a}o Paulo, CP 66318,
CEP 05315-970 S\~{a}o Paulo, SP, Brazil}
\affiliation{Department of General and Experimental Physics, Herzen State
Pedagogical University of Russia, Moyka emb. 48, 191186 St. Petersburg,
Russia}

\author{D. M. Gitman}
\email{gitman@dfn.if.usp.br}

\affiliation{Institute of Physics, University of S\~{a}o Paulo, CP 66318,
CEP 05315-970 S\~{a}o Paulo, SP, Brazil}

\author{N. Yokomizo}
\email{yokomizo@fma.if.usp.br}

\affiliation{Institute of Physics, University of S\~{a}o Paulo, CP 66318,
CEP 05315-970 S\~{a}o Paulo, SP, Brazil}

\begin{abstract}
Our previous results on the nonperturbative calculations of the mean current
and of the energy-momentum tensor in QED with the $T$-constant electric
field are generalized to arbitrary dimensions. The renormalized mean values
are found; the vacuum polarization and particle creation contributions to
these mean values are isolated in the large $T$-limit, the vacuum
polarization contributions being related to the one-loop effective
Euler-Heisenberg Lagrangian. Peculiarities in odd dimensions are considered
in detail. We adapt general results obtained in $2+1$ dimensions to the
conditions which are realized in the Dirac model for graphene. We study the
quantum electronic and energy transport in the graphene at low carrier
density and low temperatures when quantum interference effects are
important. Our description of the quantum transport in the graphene is based
on the so-called generalized Furry picture in QED where the strong external
field is taken into account nonperturbatively; this approach is not
restricted to a semiclassical approximation for carriers and does not use
any statistical assumptions inherent in the Boltzmann transport theory. In
addition, we consider the evolution of the mean electromagnetic field in the
graphene, taking into account the backreaction of the matter field to the
applied external field. We find solutions of the corresponding Dirac-Maxwell
set of equations and with their help we calculate the effective mean
electromagnetic field and effective mean values of the current and the
energy-momentum tensor. The nonlinear and linear $I-V$ characteristics
experimentally observed in both low and high mobility graphene samples is
quite well explained in the framework of the proposed approach, their
peculiarities being essentially due to the carrier creation from the vacuum
by the applied electric field.
\end{abstract}

\maketitle

\section{Introduction\label{S.1}}

\subsection{General}

It is well-known that QFT with an external background is an adequate model
for studying quantum processes in the cases when a part of the quantized
field is strong enough to be treated as a given and a classical one.
Numerous problems in QED and QCD with superstrong electromagnetic fields,
which must be treated nonperturbatively, are at present investigated in this
framework, with applications to astrophysics and condensed matter physics
(e.g. graphene physics) \cite{FGS,greiner,ruffini,allor}. In these models,
one often needs to know how the external background affects admitted states
of the quantized fields and to what extent the external background
idealization is consistent, i.e., whether the backreaction on the external
background is sufficiently small. Here we consider the quantized Dirac field
interacting with an external electromagnetic background. The external
electromagnetic field changes the states of the Dirac field and induces
transitions between them. These changes in general occur as vacuum
polarization effects combined with those due to particle creation. A uniform
constant magnetic field, for instance, can produce only vacuum polarization
effects, whereas particle creation is due to electric-like external fields.

The first calculations of vacuum polarization effects have been carried out
by Heisenberg and Euler in the case of QED with constant, uniform, parallel
electric $E$ and magnetic $B$ fields, with $E \ll E_c=m^{2}/e \simeq
10^{16}V/cm$ ($E_c$ is Schwinger's critical field) and arbitrary $B$,
ignoring particle creation \cite{HeiE36}. It turns out that the vacuum
polarization effects (which are local in time in the uniform background) can
be described by some effective Lagrangians, in particular by the
Heisenberg-Euler Lagrangian, see \cite{dunne}. On the other hand, the
particle creation effects are global ones, as they depend on the history of
the external field action and,\ therefore, cannot be described by any local
effective Lagrangian (this does not contradict the fact that the imaginary
part of the Heisenberg-Euler effective action is related to the total number
of particle created, since this action is a global quantity). A
non-perturbative approach (with respect to the external background) to the
particle creation in the so-called $t$-electric steps (electric-like fields
that are switched on and off at initial and final time instants,
respectively) was elaborated in the framework of relativistic quantum
mechanics in \cite{nikishov}, and in the framework of QFT in \cite{FGS}; a
more complete list of publications on\ the particle creation can be found in
\cite{ruffini}. A direct way to study the influence of an external
electromagnetic background on the Dirac field is calculating
nonperturbatively the mean energy-momentum tensor. In case of a
quasiconstant electric field (the so-called $T$-constant electric field%
\footnote{%
It should be noted that particle creation in the $T$-constant electric field
was first considered in Ref. \cite{BagGiS75}.}, i.e. a uniform electric
field, constant within a time interval $T$, and zero outside it) such a
problem was solved in \cite{Gav05,GG08-b}. The obtained result allows one to
isolate, under certain conditions, vacuum polarization effects from the
particle creation effects.

Until recently, problems related to particle creation from the vacuum were
of a purely theoretical interest. This is related to the fact that, due to
the presence of large gaps between the upper and lower branches in the
spectrum of Dirac particles, particle creation effects can be observed only
in huge external electric fields of the magnitude of $E_c$. However, recent
technological advances in laser science suggest that lasers such as those
planned for the Extreme Light Infrastructure project (ELI) may be able to
reach the non-perturbative regime of pair production in the near future (see
review \cite{Dun09}). Moreover, the situation has changed completely in
recent years regarding applications to condensed matter physics: particle
creation became an observable effect in graphene physics, an area which is
currently under intense development \cite{castroneto,dassarma}. Briefly,
this is explained by two facts: first, the low-energy electronic excitations
in the graphene monolayer in the presence of an external electromagnetic
field can be described by the Dirac model \cite{semenoff}, namely, by a $2+1$
quantized Dirac field in such a background (that is, dispersion surfaces are
the so-called Dirac cones); and, second, the gap between the upper and lower
branches in the corresponding Dirac particle spectra is very small, so that
the particle creation effect turns out to be dominant (under certain
conditions) as a response to the external electric-like field action on the
graphene. In particular, this effect is crucial for understanding the
conductivity of graphene, specially in the so-called non-linear regime \cite%
{allor,dora,lewkowicz-10a,lewkowicz-10b,lewkowicz-11}. The first
experimental observation of non-linear current-voltage characteristics ($I-V
$) of graphene devices and its interpretation in terms of pair-creation has
been recently reported \cite{vandecasteele}.

It was also shown recently in \cite{FefM12} that linear dispersion surfaces
with the Dirac points, similar to those in the graphene monolayers, are
generic in the spectrum of non-relativistic $2+1$-dimensional Hamiltonians
with potentials that have the symmetry of a honeycomb structure. Thus, one
may expect the discovery of new materials in which the conductivity is
described by a Dirac model of $(2+1)$ massless fermions. Moreover, a mass
gap in the graphene band structure can be generated by several methods. One
example is given by graphene nanoribbons. In this case one has a
quasi-one-dimensional spatial geometry that confines the graphene electrons
to a strip of large length and small width. The confinement gap depends on
the strip width and on imposed boundary conditions, see \cite{dassarma} for
a review. The spectrum of graphene can also be gapped by explicit or
spontaneous sublattice symmetry breaking, see \cite{GusShCar07} and
references therein. It is an important fundamental and practical problem
under current research.

It should also be noted that there is a deep connection between graphene,
which has two Dirac cones, and topological insulators, which are
characterized by a single Dirac cone on each surface; see \cite%
{dassarma,top-insul11} for a review and \cite{HgTe-exp} for the first
experimental realization of a single-valley Dirac system in zero-gap HgTe
quantum wells. In these Dirac systems, the mass gap can be generated by an
appropriate selection of material parameters that depend on the quantum well
geometry. In our consideration, it is assumed that the two cones of graphene
are decoupled and the system behaves like two copies of a single Dirac cone.
Thus, {the results obtained for graphene should also be} relevant for the
single Dirac cone on the surface of a topological insulator. Then we expect
the interface transport properties of topological insulator to be similar to
those described for graphene.

In the present article, we adapt our general results on the particle
creation to the cases which are realized in the Dirac model for graphene.
First, our previous results in calculating the vacuum mean current and the
energy-momentum tensor of the Dirac field in the $T$-constant electric field
\cite{Gav05,GG08-b} is extended to arbitrary dimensions, especially to $(2+1)
$-d. Then we consider some of immediate applications of these results to a
number of problems related to the dc conductivity of graphene in the
so-called superlinear (non-perturbative) regime.

We show that the application of non-perturbative methods of QED with strong
field and unstable vacuum to studying the Dirac model in graphene allows one
to more adequately describe quantum transport in graphene. The matter is
that usually the electronic transport in graphene, even near the Dirac
point, is described within the framework of traditional methods used in
condensed matter physics. For example, these are methods based on WKB
description of the carriers \cite{dora,vandecasteele}, or numerical methods
that attend one-particle quantum mechanical description of the carriers \cite%
{lewkowicz-10b}. Other numerical methods exploit a Green's function
formalism of non-equilibrium statistical mechanics adopted for describing
states not very far from the equilibrium \cite{vandecasteele}. In some cases
the Boltzmann transport theory, based on WKB approximation, describes the
system evolution quite well, and sometimes it allows one to find analytic
solutions. However, these cases are restricted to the ones when the large
carrier densities obey the assumption that characteristic length scales of
the system be larger than typical wave lengths of carriers. Moreover, the
Boltzmann theory is not adapted to describe motion of massless carries
created from the vacuum. The numerical methods in their present form are not
adjusted for studying the time-evolution of the system that would take the
electrodynamic backreaction into account, whereas the latter may be of
essential importance in the present highly non-equilibrium environment
caused by the pair creation. Moreover, the electric field creates carriers
in pure states, whose distribution differs significantly from the
equilibrium distribution. This is a nonstandard situation for usual
transport problems in the condensed matter physics. Thus, a proper
description of the quantum transport in graphene close to the Dirac point
both in the ballistic case and in the presence of a disorder is still an
open problem.

Our description of the quantum transport in the graphene is based on
strong-field QED, it is not restricted by a semiclassical approximation of
carriers and it does not use any statistical suppositions typical of the
Boltzmann transport theory. Our approach is based on exact solutions of the
Dirac equation, where the strong external field is taken into account
non-perturbatively. A strong field approximation used for analytical
calculations is related to the consistency of the Dirac model with a given
external field and we show that it is well-working under certain conditions.
In fact, we study the electronic and energy transport in graphene at low
carrier density and low temperatures when quantum interference effects are
important.

We consider the evolution of the mean electromagnetic field in the graphene,
taking into account the backreaction of the quantum matter field on the
applied external field. In doing this, we use consistency restrictions that
describe the regime when the backreaction can be neglected. We derive
restrictions from above on the allowed strength of the external electric
field and on its duration, admitted at its given strength. In making some
experimental conclusions, we try to compare these restrictions with typical
experimental scales. We present a generalization of the QED model with the $T
$-constant external field to take into account the backreaction of the mean
current to the applied electric field. We find a self-consistent solution of
the Dirac-Maxwell set of equations for this generalized model and calculate
the effective mean field and effective mean values of the current and the
energy-momentum tensor. We show that the non-linear and linear $I-V$
experimentally observed in low and high mobility samples, respectively, can
be explained in the framework of the presented consideration, and that such
a behavior is a consequence of the fact that the conductivity in the
graphene is essentially due to the pair creation from the vacuum by the
applied electric field.

The article is organized as follows: In Subsection \ref{SS1.2} we give
introductory overview of the basics of the QED with unstable vacuum and
introduce the necessary general notation. In Section \ref{S2}, we introduce
the Dirac model in a $T$-constant background, we fix notation and collect
previous results (mainly of \cite{GG08-b,GG95}) required for our work. Exact
solutions of the Dirac equation for this background are presented, and we
describe how they are related to the particle creation and to the mean
values of field observables, following the general theory presented in \cite%
{FGS}. In Section \ref{S3} we consider the vacuum mean values of the current
density and the energy-momentum tensor in a $T$-constant background for the
case of arbitrary dimensions. Namely, in Subsection \ref{SS3.1} we express
the vacuum mean values of the current density and the energy-momentum tensor
in terms of appropriate Green functions constructed by the help of the exact
solutions of the Dirac equation in the external field. In Subsection \ref%
{SS3.2} peculiarities in odd dimensions are considered. In Section \ref%
{SS3.3} vacuum polarization and particle creation contributions to the mean
values of the energy-momentum tensor are\emph{\ }isolated in the large
duration approximation, and the vacuum polarization contributions are
related to the one-loop effective Euler-Heisenberg Lagrangian. In Subsection %
\ref{SS3.4} we find renormalized mean values of the energy-momentum tensor
using the zeta-function regularization and compute the vacuum polarization
contributions to these mean values in arbitrary dimensions. In Subsection %
\ref{SS3.5} we compute the particle creation contributions to the mean
values of the current and of the energy-momentum tensor in arbitrary
dimensions\ to generalize the four dimension results of \cite{GG08-b}.
Pair-creation contributions are finite, due to the finite duration of the
field. In Section \ref{S4}, our results are applied to graphene physics.
Namely, in Subsection \ref{SS4.1} the Dirac model of graphene is briefly
described, and the mean values computed in the previous section are used to
determine the mean current and energy-momentum tensor in the material. In
Subsection  \ref{SS4.2} these results allow us to estimate scales for
backreaction effects, using the appropriate solution of the Maxwell
equations. We show that the superlinearity of $I-V$\ observed experimentally
in low mobility samples is essentially due to the pair creation from the
vacuum by the applied electric field in the regime where \ the backreaction
is negligible. In Subsection  \ref{SS4.3} we study a generalization of the
model with the $T$-constant external field to taking into account the
backreaction of the mean current to the applied electric field. We find a
self-consistent solution of the Dirac-Maxwell set of equations (with
unstable vacuum in fermion sector) and calculate the effective mean field
and effective mean values of the current and of energy-momentum tensor. We
show that linear $I-V$\ experimentally observed in high mobility samples is
due to the pair creation from the vacuum by the applied electric field in
the regime where backreaction is important. In Section \ref{S5} (Summary) we
briefly list the main new results obtained in the article and add some
relevant comments.

\subsection{Dirac model with external background\label{SS1.2}}

In order to study the Dirac field in the external background, we use a
formulation of QED with unstable vacuum (the generalized Furry
representation) developed in \cite{FGS,GavGT06}, where a strong external
field is treated nonperturbatively. In this work we consider the case of the
so-called $T$-constant electric field, i.e., we assume that for $t<t_{in}$
and for $t>t_{out}$, the $T$-constant electric field is absent, therefore
initial and final vacua are vacuum state of free $in$- and $out$- particles
respectively. During the interval $t_{out}$ $-t_{in}$ $=T$, the Dirac field
interacts with a constant uniform electric field. The initial and final
vacua are different due to the difference of the initial and final value of
the external electromagnetic field potentials. In the Heisenberg picture,
there exists a set of creation and annihilation operators $a_{n}^{\dagger
}(in)$, $a_{n}(in)$ of $in$-particles (electrons), and operators $%
b_{n}^{\dagger }(in)$, $b_{n}(in)$ of $in$-antiparticles (positrons), the
corresponding $in$-vacuum being $|0,in\rangle$ at the same time there exists
a set of creation and annihilation operators $a_{n}^{\dagger }(out)$, $%
a_{n}(out)$ of $out$-electrons and operators $b_{n}^{\dagger }(out)$, $%
b_{n}(out)$ of $out$-positrons, the corresponding $out$-vacuum {being} $%
|0,out\rangle $,
\begin{align*}
& a_{n}(in)|0,in\rangle =b_{n}(in)|0,in\rangle =0, \; \forall n, \\
& a_{n}(out)|0,out\rangle =b_{n}(out)|0,out\rangle =0, \; \forall n.
\end{align*}
In both cases, by $n$ we denote complete sets of quantum numbers describing $%
in$- and $out$- particles. The $in$- and $out$-operators obey the canonical
anticommutation relations:
\begin{align*}
& \lbrack a_{n}(in),a_{n^{\prime }}^{\dagger}(in)]_{+} =
[a_{n}(out),a_{n^{\prime }}^{\dagger}(out)]_{+} = \delta _{n,n^{\prime }}\,,
\\
& [b_{n}(in),b_{n^{\prime }}^{\dagger}(in)]_{+} = [b_{n}(out),b_{n^{\prime
}}^{\dagger }(out)]_{+}=\delta _{n,n^{\prime }}\,,
\end{align*}
the remaining anticommutators being zero.

The $in$-operators are associated with a complete orthonormal set of
solutions $\left\{ _{\zeta }\psi _{n}(x)\right\} $ ($\zeta =+$ for electrons
and $\zeta =-$ for positrons) of the Dirac equation with $T$-constant
electric field. Their asymptotics as $t<t_{in}$ can be classified as free
particles and antiparticles. The $out$-operators are associated with a
complete orthonormal $out$-set of solutions $\left\{ ^{\zeta }\psi
_{n}\left( x\right) \right\} $ of the Dirac equation with $T$-constant
electric field. Their asymptotics as $t>t_{out}$ can be classified as free
particles and antiparticles. The $in$- and $out$- operators are defined by
the two representations of the quantum Dirac field $\Psi (x)$ in the
Heisenberg representation
\begin{align}
\Psi (x) & =\sum_{n}\left[ a_{n}(in)\;_{+}\psi _{n}(x)+b_{n}^{\dagger
}(in)\;_{-}\psi _{n}(x)\right]  \notag \\
& =\sum_{n}\left[ a_{n}(out)\;^{+}\psi _{n}\left( x\right)
+b_{n}^{\dagger}(out))\;^{-}\psi _{n}\left( x\right) \right] .
\label{in-out-repr}
\end{align}%
$In$- and $out$-solutions with given quantum numbers $n$ are related by a
linear transformation of the form:
\begin{equation}
^{\zeta }\psi _{n}\left( x\right) =g(_{+}\mid ^{\zeta })\,_{+}\psi
_{n}\left( x\right) +g(_{-}\mid ^{\zeta })\,_{-}\psi _{n}\left( x\right) \,,
\label{eq:in-out-g}
\end{equation}
where the $g^{\prime}$s are some complex coefficients. Then a linear
canonical transformation (Bogolyubov transformation) between $in$- and $out$%
- operators which follows from Eq.~(\ref{in-out-repr}) is defined by these
coefficients.

The vacuum mean current vector, energy and momentum vector of the quantum
Dirac field $\Psi(x)$ at a time instant $t$ are defined as integrals over
the spatial volume. Due to translational invariance of the external field
under consideration, all these mean values are proportional to the space
volume. Therefore, it is enough to calculate the vacuum mean values of the
current density vector $\langle j^{\mu }(t)\rangle$ and of the
energy-momentum tensor (EMT) $\langle T_{\mu \nu }(t)\rangle$,
\begin{eqnarray}
&\langle j^{\mu }(t)\rangle =&\langle 0,in|j^{\mu }|0,in\rangle \,,  \notag
\\
&\langle T_{\mu \nu }(t)\rangle =&\langle 0,in|T_{\mu \nu }|0,in\rangle \,.
\label{int1}
\end{eqnarray}
Here we stress the time-dependence of mean values (\ref{int1}), which does
exist due to the time-dependence of the external field. The operators of the
current density and the EMT are described in terms of the quantum Dirac
field as follows\footnote{%
We use the relativistic units $\hslash =c=1$, in which the fine structure
constant is $\alpha =e^{2}/c\hslash =e^{2}$.}
\begin{align*}
& j^{\mu }=\frac{q}{2}\left[ \bar{\Psi}(x),\gamma ^{\mu }\Psi (x)\right]
\,,\quad T_{\mu \nu }=\frac{1}{2}(T_{\mu \nu }^{can}+T_{\nu \mu }^{can})\,,
\\
& T_{\mu \nu }^{can}=\frac{1}{4}\left\{ [\bar{\Psi}(x),\gamma _{\mu }P_{\nu}
\Psi (x)]+[P_{\nu }^{\ast }\bar{\Psi}(x),\gamma _{\mu }\Psi (x)]\right\} \,,
\\
& P_{\mu }=i\partial _{\mu }-qA_{\mu }(x),\ \bar{\Psi}(x)=\Psi ^{\dagger}
(x)\gamma ^{0},
\end{align*}
where $A_{\mu }(x)$ are electromagnetic potentials of the external field, $q$
is the particle charge (for an electron $q=-e$), and $\Psi (x)$ is the
Heisenberg operator of the Dirac field that obeys the Dirac equation with
the external background.

Note that the mean values (\ref{int1}) depend on the definition of the
initial vacuum, $|0,in\rangle $ and on the evolution of the electric field
from the time $t_{in}$ of switching it on up to the current time instant $t$%
, but they have nothing to do with the further history of the system. The
renormalized vacuum mean values $\langle j^{\mu }(t)\rangle $ and $\langle
T_{\mu \nu }(t)\rangle ,$ $t_{in}$ $<t<$ $t_{out}$ are sources in equations
of motion for mean electromagnetic and metric fields, respectively. In
particular, complete description of the backreaction is related to the
calculation of these mean values for any $t$.

\section{Dirac field in $T$-constant electric background\label{S2}}

In what follows we are going to deal with the Dirac equation in $(d=D+1)$%
-dimensional Minkowski space with an external electromagnetic field $A_{\mu}
(x)$,
\begin{equation}
(\gamma^\mu P_\mu - m) \psi(x) = 0 \, .  \label{eq:dirac-equation}
\end{equation}%
Here $\psi(x)$ is a $2^{[d/2]}$-component spinor (the brackets stand for
`integer part of'), $m$ is the mass, $q$ is the charge, the Greek index
assumes values $\mu =0,1,\dots ,D$, and the gamma matrices satisfy the
standard anti-commutation relations:
\begin{equation*}
\left[ \gamma ^{\mu },\gamma ^{\nu }\right] _{+}=2\eta ^{\mu \nu },\ \ \eta
_{\mu \nu }= \mathrm{diag}(1,-1,\ldots ,-1).
\end{equation*}

In what follows, we consider the $T$-constant field described by a vector
potential with only one nonzero component $A_{1}(t)$ ($A_{\mu}(t)=0,\ \mu
\neq 1$),
\begin{equation*}
A_{1}(t)=E\left\{
\begin{array}{ll}
t_{in} & t\in \mathrm{I}=(-\infty ,t_{in}),\ t_{in}=-T/2 \, , \\
t, & t\in \mathrm{Int}=[t_{in},t_{out}] \, , \\
t_{out}, & t\in \mathrm{II}=(t_{out},\infty ) \, ,\ t_{out}=T/2 \, .%
\end{array}%
\right.
\end{equation*}%
such that the electric field $\mathbf{E}\left( t\right)$ has also only one
nonzero component, which is nonzero for $t\in \mathrm{Int}$, i.e.,
\begin{equation*}
E^{1}\left( t\right) =E,\; t\in \mathrm{Int}; \quad E^{1}(t)=0 \, , \; t\in
\mathrm{I\cup II}.
\end{equation*}

If one represents the spinor $\psi (x)$ in the form
\begin{equation}
\psi (x)=(\gamma ^{\mu }P_{\mu }+m)\phi (x)\,,  \label{eq:psi-phi}
\end{equation}%
where $\phi (x)$ is a new spinor, then $\phi (x)$ obeys the following
equation:
\begin{equation}
\left( P^{2}-m^{2}-\frac{q}{2}\sigma ^{\mu \nu }F_{\mu \nu }\right) \phi
(x)=0,  \label{eq:dirac-phi}
\end{equation}%
where%
\begin{align*}
& F_{\mu \nu }=\partial _{\mu }A_{\nu }-\partial _{\nu }A_{\mu },\ \ \sigma
^{\mu \nu }=\frac{i}{2}[\gamma ^{\mu },\gamma ^{\nu }] \, , \\
& \frac{q}{2}\sigma ^{\mu \nu }F_{\mu \nu }=iqE(t)\gamma ^{0}\gamma ^{1} \, .
\end{align*}
Solutions of the Dirac equation in a $T$-constant field were studied in
detail in \cite{GG95}. Below, we use these results.

First we choose a set of constant orthonormalized spinors $v_{s,r}$, with $%
s=\pm 1$, and $r=(r_{1},r_{2},\dots ,r_{[d/2]-1})$, $r_{i}=\pm 1$, such that
$\gamma ^{0}\gamma ^{1}v_{s,r}=sv_{s,r}$. The indices $r_{i}$ describe the
spin polarization, which is not coupled to the electric field, and together
with the additional index $s$ provide a suitable parametrization of the
solutions. Then we represent the spinors $\phi (x)$ as follows:
\begin{equation}
\phi _{\mathbf{p,}s,r}(t,\mathbf{x})=e^{i\mathbf{p}\cdot \mathbf{x}}\varphi
_{\mathbf{p,}s}(t)v_{s,r} \,.  \label{eq:phi-varphi}
\end{equation}%
Thus, the time-evolution is described by the ordinary differential equation
of second order
\begin{equation}
\left\{ \frac{d^{2}}{dt^{2}}+\left[ p_{1}-qA_{1}(t)\right] ^{2}+\left\vert
qE\right\vert \lambda +isqE(t)\right\} \varphi _{\mathbf{p,}s}(t)=0\,,
\label{eq:time-evolution}
\end{equation}%
where $\lambda =(\mathbf{p}_{\perp }^{2}+m^{2})/\left\vert qE\right\vert $
and $\mathbf{p}_{\perp }$ is the transversal momentum, $\mathbf{p}_{\perp
}=(0,p^{2},\dots ,p^{D})$.

At early ($t<t_{in}$ -region $\mathrm{I}$), and late ($t>t_{out}$ -region $%
\mathrm{II}$) times, Eq.~(\ref{eq:time-evolution}) has plane wave solutions $%
_{\zeta }\phi _{\mathbf{p,}s,r}$ and $^{\zeta }\phi _{\mathbf{p,}s,r}$
respectively, with $\zeta =\pm 1$, which satisfy simple dispersion relations%
\begin{align}
&\mathrm{I}:\ _{\zeta }\varphi _{\mathbf{p},s}(t)\sim e^{-i\zeta
\omega_{in}t} \, , \quad \mathrm{II}:\ ^{\zeta }\varphi _{\mathbf{p}%
,s}(t)\sim e^{-i\zeta\omega _{out}t} \, ,  \notag \\
&\omega _{in/out}=\sqrt{(p_{1}-qEt_{in/out})^{2}+\mathbf{p}_{\perp
}^{2}+m^{2}} \, .  \label{free-solutions}
\end{align}
As was demonstrated in \cite{GG95}, the $in$-set $\{_{\zeta }\psi _{\mathbf{%
p,}r}\}$ and $out$-set $\{^{\zeta }\psi _{\mathbf{p,}r}\}$ of solutions of
the Dirac equation in the $T$-constant electric field can be taken in the
form
\begin{gather}
_{\pm }\psi _{\mathbf{p,}r}(x)=(\gamma \cdot P+m)_{\pm }\phi _{\mathbf{p}%
,\pm ,r}(x)\,,  \notag \\
^{\pm }\psi _{\mathbf{p,}r}(x)=(\gamma \cdot P+m)^{\pm }\phi _{\mathbf{p}%
,\mp ,r}(x)\,.  \label{eq:in-out-solutions}
\end{gather}

For $t\in \mathrm{Int}$, the general solution of Eq.~%
\eqref{eq:time-evolution} is completely determined by an appropriate pair of
the following linear independent Weber parabolic cylinder functions:
\begin{equation*}
D_{\nu -(1+s) /2}\left( \zeta (1-i)\xi \right) ,\ \ D_{-\nu -(1-s) /2}\left(
\zeta (1+i)\xi \right) \, ,
\end{equation*}
where $\zeta =\pm$, $\nu=i\lambda/2$, and
\begin{equation*}
\xi =\xi (t)=\frac{\left\vert qE\right\vert t-p_{1}\mathrm{sgn}\left(
qE\right) }{\sqrt{\left\vert qE\right\vert }}\,.
\end{equation*}
According to relation (\ref{eq:in-out-g}), an $out$-solution corresponding
to a plane wave in interval $\mathrm{II}$ is thus described by a
superposition of the Weber functions in {interval} $\mathrm{Int}$, and
extends into a superposition of particle and antiparticle $in$-solutions in $%
\mathrm{I}$. Then one can explicitly find the coefficients $g$. In terms of
such coefficients, the differential mean number $\aleph _{\mathbf{p},r}$ of
particles created from vacuum with given $\mathbf{p}$ and $r$ at a time
instant $t>t_{out}$ is
\begin{equation}
\aleph _{\mathbf{p},r}=\langle 0,in|a_{\mathbf{p},r}^{\dagger }(out)a_{%
\mathbf{p},r}(out)|0,in\rangle =\aleph _{\mathbf{p}}=|g(_{-}\mid
^{+})|^{2}\,,  \label{eq:np-formula}
\end{equation}
where the Bogolyubov transformation between $in$- and $out$- creation and
annihilation operators is used. The number of particles created is equal to
the numbers of antiparticles created. Then $\aleph_{\mathbf{p},r}$ can be
treated as the number of pairs created. The result in (\ref{eq:np-formula})
is independent of the spin polarization. That is why we use in what follows
the notation $\aleph _{\mathbf{p}}$ for the quantity (\ref{eq:np-formula}).
Note that there is no particle production after the time instant $t_{out}$.
Thus, $\aleph _{\mathbf{p}}$ depends only on the interval $T$. The explicit
expression for $\aleph _{\mathbf{p}}$ was studied in detail in \cite{GG95}.
Here we just quote the relevant results for the calculation of the vacuum
mean values of the EMT and current vector.

The electric field acting during the time $T$ creates a considerable number
of pairs only in a finite region in the momentum space. One can introduce a
cutoff $K>>\max \left\{ 1,m^{2}/\left\vert qE\right\vert \right\} $ such
that, for $\sqrt{\left\vert qE\right\vert }T/2>K$, one needs to consider
only the region
\begin{gather}
\left\vert \mathbf{p}_{\bot }\right\vert \leq \sqrt{\left\vert qE\right\vert
}\left[ \sqrt{\left\vert qE\right\vert }T-K\right] ^{1/2} \, ,  \notag \\
-T/2+K/\sqrt{\left\vert qE\right\vert }\leq p_{1}/qE\leq T/2-K/\sqrt{%
\left\vert qE\right\vert }\,  \label{T-cutoff}
\end{gather}
in the momentum space. After the cutoff, physical quantities that are
expressed in terms of $\aleph _{\mathbf{p}}$ will, in general, depend on $K$%
. But if $T$ is big enough, so that
\begin{equation}
\sqrt{\left\vert qE\right\vert }T \gg K \gg \max \left\{ 1,m^{2}/\left\vert
qE\right\vert \right\} ,  \label{stabilization-condition}
\end{equation}
the dependence on $K$ can be ignored. We shall suppose that the
stabilization condition (\ref{stabilization-condition}) holds true. In this
case the differential mean numbers of created pairs have the form
\begin{equation}
\aleph _{\mathbf{p}}\simeq e^{-\pi \lambda } \, ,  \label{eq:Np-asymptotic}
\end{equation}%
which is the same for the case of the constant uniform electric field.

Taking into account cutoff (\ref{T-cutoff}), we find that the total number $%
\aleph $ of pairs created is proportional to $d-1$-dimensional spatial
volume $V_{(d-1)}$ and can be expressed through the total number density $%
n^{cr}$ of pairs created during the interval $T$ as follows:
\begin{align}
& \aleph =\frac{V_{\left( d-1\right) }}{\left( 2\pi \right) ^{d-1}}\int d%
\mathbf{p}\sum_{r}\aleph _{\mathbf{p}}=V_{\left( d-1\right) }n^{cr},  \notag
\\
& n^{cr}=r^{cr}\left[ T+\left\vert qE\right\vert ^{-1/2}O\left( K\right) %
\right] \, ,  \notag \\
& r^{cr}=\frac{2^{[d/2]-1}}{(2\pi )^{d-1}}\left\vert qE\right\vert
^{d/2}\exp \left( -\frac{\pi m^{2}}{\left\vert qE\right\vert }\right) \, ,
\label{n-cr}
\end{align}
where the quantity $r^{cr}$ is often called the pairs production rate.

It should be noted that the stabilization condition is written for arbitrary
external field $E$. However, in what follows we are interested in strong
electric fields%
\begin{equation}
E>E_{c}\Longrightarrow m^{2}/\left\vert qE\right\vert <1.
\label{strong field}
\end{equation}%
In what follows, when ever we speak of the strong field case we will have in
mind the condition (\ref{strong field}). In these cases the stabilization
condition is simplified,%
\begin{equation}
\sqrt{\left\vert qE\right\vert }T>>K>>1.  \label{stab1}
\end{equation}%
The stabilization condition reveals another important dimensionless
parameter $\tau$,
\begin{equation}
\tau =\sqrt{\left\vert qE\right\vert }T\ .  \label{tau}
\end{equation}%
In the strong $T$-constant field under consideration the qualitative
supposition that $T$ is big enough is equivalent to the supposition that $%
\tau \gg 1$. In what follows, we calculate different mean values in an
approximation that is related to large $\tau$; we call such an approximation
large $\tau$-limit. To explain the meaning of such a limit, we consider
possible structures of the mean values that appear in our further
calculations. The most general structure of the mean values has the form%
\begin{equation}
\langle F\rangle =\sum_{n=1}^{\infty }F_{-n}\tau ^{-n}+F_{0}+\tilde{F}\ln
\tau +F_{1}\tau +F_{2}\tau ^{2}.  \label{GenStr}
\end{equation}%
The large $\tau $-limit for the mean value $\langle F\rangle $ means the
leading term approximation for (\ref{GenStr}), having in mind the hierarchy $%
F_{0}\ll \tilde{F}\ln \tau \ll F_{1}\tau \ll F_{2}\tau ^{2}.$ For example,
the r.h.s. in Eq.~(\ref{eq:Np-asymptotic}) represents the term $F_{0},$
while $r^{cr}$ in Eq.~(\ref{n-cr}) represents $F_{1}$ in the general formula
(\ref{GenStr}).

It should be noted that below we encounter cases where the large $\tau $%
-limit is defined by $\tau =\sqrt{\left\vert qE\right\vert }\Delta t,$ with
some $\Delta t,$ such that $\tau \gg K \gg 1$.

It should also be noted that for large $T$, the $in$- and $out$-solutions in
the interval $t\in \mathrm{Int}$ take the form:
\begin{gather}
_{+}^{-}\varphi _{\mathbf{p,}s}(t)=CD_{\nu -\left( 1+s\right) /2}\left( \pm
(1-i)\xi \right) \,,  \notag \\
_{-}^{+}\varphi _{\mathbf{p,}s}(t)=CD_{-\nu -\left( 1-s\right) /2}\left( \pm
(1+i)\xi \right) \,,  \label{eq:varphi-modes}
\end{gather}%
with the normalization constant
\begin{equation}
C=(2\pi )^{-\left( d-1\right) /2}\left\vert 2qE\right\vert ^{-1/2}\exp (-\pi
\lambda /8)\,.  \label{eq:C-def}
\end{equation}

It is supposed that the measurement is carried out at some time after
switching off the electric field, i.e., decoherence occurs after the
electric field is switched off. Of course, one can consider the case when
decoherence occurs earlier, for example, at an instant $t_{dec}$, $t_{in}$ $%
<t_{dec}<t_{out}$ . Let us suppose, for example, that the interval $%
t_{dec}-t_{in}$ satisfies stabilization conditions similar to (\ref%
{stabilization-condition}), $\sqrt{\left\vert qE\right\vert }\left(
t_{dec}-t_{in}\right) \gg K$. Then the differential mean number of pairs
created in the large $t_{dec}-t_{in}$ limit, when%
\begin{gather}
\left\vert \mathbf{p}_{\bot }\right\vert \leq \sqrt{\left\vert qE\right\vert
}\left[ \sqrt{\left\vert qE\right\vert }\left( t_{dec}-t_{in}\right) -K%
\right] ^{1/2}\;,  \notag \\
t_{in}+K/\sqrt{\left\vert qE\right\vert }\leq p_{1}/qE\leq t_{dec}-K/\sqrt{%
\left\vert qE\right\vert }\,,  \label{dec-cutoff}
\end{gather}%
is given by (\ref{eq:Np-asymptotic}) and the total number density of
particles created is $r^{cr}\left( t_{dec}-t_{in}\right) $. For the time of
decoherence $t_{dec}$, which is sufficiently close to $t_{out}$ , $\left(
t_{out}-t_{dec}\right) /T\ll 1$, the difference of this density from the
density $n^{cr}$ can be ignored and the interpretation of particles at $%
t_{dec}$ as final out-particles already makes sense.

However, if the interval $t_{out}$ $-t_{dec}$ is comparable with the
interval $t_{dec}-t_{in}$ then it is necessary to consider the further
evolution of many-particle state given initially by the distribution (\ref%
{eq:Np-asymptotic}) at the time $t_{dec}$, which is a completely different
task, see for example \cite{GavGT06}. In this case, the differential mean
numbers of additional pairs created by the external field during the
interval from $t_{dec}$ to $t_{out}$ is given by Eq.~(59) of the work \cite%
{GavGT06} as follows:
\begin{equation}
\Delta N_{\mathbf{p}}=\aleph _{\mathbf{p}}\left[ 1-\left( N_{\mathbf{p}%
}^{(+)}(in)+N_{\mathbf{p}}^{(-)}(in)\right) \right] ,  \label{delta-N}
\end{equation}%
where $N_{\mathbf{p}}^{(+)}(in)=N_{\mathbf{p}}^{(-)}(in)=\aleph _{\mathbf{p}%
} $ are differential mean numbers of initial particles and antiparticles at $%
t_{dec}$, given by Eq.~(\ref{eq:Np-asymptotic}). If $N_{\mathbf{p}%
}^{(+)}(in)+N_{\mathbf{p}}^{(-)}(in)>1$ then $\Delta N_{\mathbf{p}}$ is
negative, that is, the annihilation of the existing pairs of particles and
antiparticles results from the electric field. Integrating and summing
distribution (\ref{delta-N}) over the quantum numbers under conditions (\ref%
{dec-cutoff}), we obtain the additional number density of pairs
\begin{equation}
\Delta n^{cr}=\left[ 1-2^{2-d/2}\exp \left( -\frac{\pi m^{2}}{\left\vert
qE\right\vert }\right) \right] r^{cr}\left( t_{dec}-t_{in}\right) .
\label{delta-n}
\end{equation}%
Summing the density of pairs created at $t_{dec}$, $r^{cr}\left(
t_{dec}-t_{in}\right) $, the additional density $\Delta n^{cr}$, and the
number density of pairs created from vacuum states after $t_{dec}$, $%
r^{cr}\left( t_{out}-t_{dec}\right) $, we find the following total number
density of pairs created during the interval $T,$
\begin{equation*}
n^{cr}\left( t_{out}-t_{dec}\left\vert t_{dec}-t_{in}\right. \right)
=n^{cr}+\Delta n^{cr}.
\end{equation*}%
Thus, if decoherence occurs at $t_{dec}$, the final number density of pairs
at the time instant $t_{out}$ differs from $n^{cr}$ by $\Delta n^{cr}$. In
the case of strong electric field, $\frac{m^{2}}{\left\vert qE\right\vert }%
\ll 1$, $\Delta n^{cr}\approx 0$ for $d=4$, $\Delta n^{cr}<0$ for $d=2,3$,
and $\Delta n^{cr}>0$ for $d>4$. We see that the intermediate decoherence at
the early time significantly reduces the measured density of final particles
in low-dimensional systems, and increases in high-dimensional systems.
Decoherence can occur once or many times during the interval $T$. The choice
of a suitable model of decoherence depends on the physical nature of the
phenomenon.

It is useful to comment the case when the initial many-particle state is the
thermodynamical equilibrium of noninteracting particles at the temperature $%
\Theta $ with the chemical potential $\mu _{ch}$,%
\begin{equation*}
N_{\mathbf{p}}^{(\pm )}(in)=\left\{ \exp \left[ \left( \omega _{in}-\mu
_{ch}\right) /\Theta \right] +1\right\} ^{-1},
\end{equation*}%
where $\omega _{in}$ is given by Eq.~(\ref{free-solutions}). Then the
differential mean number $\Delta N_{\mathbf{p}}$ of additional pairs created
by the external field during the interval $T$, has the form (\ref{delta-N}),
where $\aleph _{\mathbf{p}}$ is given by (\ref{eq:Np-asymptotic}). Taking
into account the cutoff (\ref{T-cutoff}), one can find the total number of
additional particles created. In consequence, one can see that at low
temperatures, $\left( m-\mu _{ch}\right) /\Theta \gg 1$, such a total number
differs from the zero-temperature result (\ref{n-cr}) by a next-to-leading
term that is not essential in the large $\tau $-limit, see \cite{GavGT06}.
We consider the system at high temperatures when all the energies of the
created and accelerated particles are much lower than the temperature $%
\Theta $, hence $\left\vert qE\right\vert T/\Theta \ll 1$. One can extract
from results of the work \cite{GavGT06}, that the total number of additional
particles created under such a condition is much less than that in (\ref%
{n-cr}). However, we stress that at high temperatures, the polarization
effect in the current and energy densities of the initial gas of charged
particles given by Eq.~(127) in \cite{GG08-b}, is much stronger than effect
from the pair creation given by Eq.~(72) in \cite{GG08-b}. This could lead
to screening the electric field before the effects of pair creation manifest
themselves. One can see that the vacuum contributions dominate in comparison
with contributions due to the low temperature and particle density in the
initial state. That is why we restrict ourselves to the case of the vacuum
initial state.

\section{Energy-momentum tensor and current vector\label{S3}}

\subsection{Observables and Green functions\label{SS3.1}}

We consider here various singular functions of the Dirac field. With the
help of these functions, different physical quantities can be calculated.
For example, mean values (\ref{int1}) can be calculated with the help of the
so-called $in$-propagator:
\begin{equation*}
S_{in}^{c}(x,x^{\prime })=i\langle 0,in|T\Psi (x)\bar{\Psi}(x^{\prime
})|0,in\rangle \,.
\end{equation*}
In turn, this propagator can be determined via the $in$-solutions as follows
\begin{align}
S_{in}^{c}(x,x^{\prime })& =\theta (t-t^{\prime })S_{in}^{-}(x,x^{\prime
})-\theta (t^{\prime }-t)S_{in}^{+}(x,x^{\prime })\,,  \notag \\
S_{in}^{\mp }(x,x^{\prime })& =i\int d\mathbf{p}\sum_{r}\,{}_{\pm }\psi _{%
\mathbf{p,}r}(x)\,_{\pm }\bar{\psi}_{\mathbf{p,}r}(x^{\prime })\,.
\label{in-in}
\end{align}%
Using relation (\ref{eq:in-out-g}) and properties of the $g$ coefficients,
one can divide the propagator (\ref{in-in}) into a sum of two terms,
\begin{equation}
S_{in}^{c}(x,x^{\prime })=S^{c}(x,x^{\prime })+S^{p}(x,x^{\prime })\,,
\label{in-in-c}
\end{equation}%
where the first term is the causal (Feynman) propagator,
\begin{align}
& S^{c}(x,x^{\prime })=i\langle 0,out|T\Psi (x)\bar{\Psi}(x^{\prime
})|0,in\rangle c_{v}^{-1} \, ,  \notag \\
& c_{v}=\langle 0,out|0,in\rangle \, .  \label{eq:causal-propagator}
\end{align}
Here $c_{v}$ is the vacuum-to-vacuum probability amplitude. This propagator
can be represented as follows:
\begin{align}
& S^{c}(x,x^{\prime}) =\theta (t-t^{\prime }) \, S^- \left( x,x^{\prime
}\right) -\theta (t^{\prime }-t) \, S^+ \left(x,x^{\prime }\right) \,,
\notag \\
& S^{-}(x,x^{\prime}) =i\int d\mathbf{p}\sum_{r}{}^{+}\psi _{\mathbf{p,}%
r}\left( x\right) g\left(_{+}|^{+}\right) ^{-1}\,_{+}\bar{\psi}_{\mathbf{p}%
,r}\left( x^{\prime }\right) \,,  \notag \\
& S^{+}(x,x^{\prime }) =i\int d\mathbf{p}\sum_{r} {}_{-}\psi _{\mathbf{p}%
,r}(x) \left[ g\left(_{-}|^{-}\right)^{-1}\right]^{\ast}{}^{-}\bar{\psi}_{%
\mathbf{p,}r}(x^{\prime}) .  \label{emt1}
\end{align}

The second term in (\ref{in-in-c}) has the following form:
\begin{multline}
S^{p}(x,x^{\prime })=i\int d\mathbf{p}\sum_{r}{}_{-}\psi _{\mathbf{p},r}(x)
\\
\times \left[ g(_{+}|^{-})g(_{-}|^{-})^{-1}\right] ^{\ast}{}_{+}\bar{\psi}_{%
\mathbf{p,}r}(x^{\prime }) \, .  \label{eq:pair-production-propagator}
\end{multline}

The current density, $j_{cr}^{\mu }(t)$, and EMT, $T_{\mu \nu }^{cr}(t)$, of
created particles at $t \agt t_{out}$, are expressed via the mean values of
the normal form of $j^{\mu }$ and $T_{\mu \nu }$ operators with respect to
the $out$-vacuum. Namely,
\begin{align}
& j_{cr}^{\mu }(t)=\langle j^{\mu }(t)\rangle -\langle j^{\mu }(t)\rangle
_{out} \, ,  \notag \\
& \langle j^{\mu }(t)\rangle _{out}=\langle 0,out|j^{\mu }|0,out\rangle \, ,
\notag \\
& T_{\mu \nu }^{cr}(t)=\langle T_{\mu \nu }(t)\rangle -\langle T_{\mu \nu
}(t)\rangle _{out} \, ,  \notag \\
& \langle T_{\mu \nu }(t)\rangle _{out}=\langle 0,out|T_{\mu \nu
}|0,out\rangle \,.  \label{cr-defc}
\end{align}%
These mean values can be calculated with the help of the so-called
out-propagator%
\begin{equation*}
S_{out}^{c}(x,x^{\prime })=i\langle 0,out|T\Psi (x)\bar{\Psi}(x^{\prime
})|0,out\rangle \,.
\end{equation*}%
Similarly to the case of the in-propagator, one can relate this propagator
to $S^{c}(x,x^{\prime })$ as follows:
\begin{equation}
S_{out}^{c}(x,x^{\prime })=S^{c}(x,x^{\prime })+S^{\bar{p}}(x,x^{\prime })
\, ,
\end{equation}
where
\begin{multline}
S^{\bar{p}}(x,x^{\prime }) =-i \int d\mathbf{p}\sum_{r}\,^{+}{\psi }_{%
\mathbf{p,}r}(x) \\
\times g(_{+}|^{+})^{-1} g(_{+}|^{-})\;{^{-}\bar{\psi}}_{\mathbf{p,}%
r}(x^{\prime }) \, .  \label{out-out}
\end{multline}

The quantities (\ref{int1}) and (\ref{cr-defc}) are real valued and can be
represented as:
\begin{align}
& \langle j^{\mu }(t)\rangle =\re \,\langle j^{\mu }(t)\rangle ^{c}+\re %
\,\langle j^{\mu }(t)\rangle ^{p} \, ,  \notag \\
& \langle T_{\mu \nu }(t)\rangle = \re \,\langle T_{\mu \nu }(t)\rangle ^{c}+%
\re \,\langle T_{\mu \nu }(t)\rangle ^{p} \, ,  \label{eq:energy-mean} \\
& \langle j^{\mu }(t)\rangle _{out}=\re\,\langle j^{\mu }(t)\rangle ^{c}+\re%
\,\langle j^{\mu }(t)\rangle ^{\bar{p}}\,,  \notag \\
& \langle T_{\mu \nu }(t)\rangle _{out}=\re \,\langle T_{\mu \nu }(t)\rangle
^{c}+\re \,\langle T_{\mu \nu }(t)\rangle ^{\bar{p}}.  \label{out-mean}
\end{align}
where
\begin{align}
& \langle j^{\mu }(t)\rangle ^{c,p,\bar{p}}=iq\left. \mathrm{tr}\left[
\gamma ^{\mu }S^{c,p,\bar{p}}(x,x^{\prime })\right] \right\vert
_{x=x^{\prime }}\,,  \notag \\
& \langle T_{\mu \nu }(t)\rangle ^{c,p,\bar{p}}=i\left. \mathrm{tr}\left[
A_{\mu \nu }S^{c,p,\bar{p}}(x,x^{\prime })\right] \right\vert _{x=x^{\prime
}}\,,  \notag \\
& A_{\mu \nu }=1/4\left[ \gamma _{\mu }\left( P_{\nu }+P_{\nu }^{\prime \ast
}\right) +\gamma _{\nu }\left( P_{\mu }+P_{\mu }^{\prime \ast }\right) %
\right] \,.  \label{eq:A-def}
\end{align}%
Here $\mathrm{tr}$ stands for the trace in the $\gamma$-matrices indices and
the limit $x\rightarrow x^{\prime }$ is understood as follows:
\begin{multline*}
\mathrm{tr} [\cdots (x,x^{\prime })]_{x=x^{\prime }} =\frac{1}{2} \left[
\lim_{t\rightarrow t^{\prime }-0} \mathrm{tr} [ \cdots (x,x^{\prime })]
\right. \\
\left. + \lim_{t\rightarrow t^{\prime }+0}\mathrm{tr}[\cdots
(x,x^{\prime})]]_{\mathbf{x=x}^{\prime }} \right] \, .
\end{multline*}%
The representations (\ref{eq:energy-mean}) and (\ref{out-mean}) imply that
\begin{align}
& j_{cr}^{\mu }(t)=\re \,\langle j^{\mu }(t)\rangle ^{p}-\re \,\langle
j^{\mu }(t)\rangle ^{\bar{p}} \, ,  \notag \\
& T_{\mu \nu }^{cr}(t)=\re \,\langle T_{\mu \nu }(t)\rangle ^{p}\,- \re %
\,\langle T_{\mu \nu }(t)\rangle ^{\bar{p}} \, .  \label{mean-cr1}
\end{align}

Note that the mean current $\,\langle j^{\mu }(t)\rangle$ and the physical
part of the mean value $\langle T_{\mu \nu }(t)\rangle$ are zero for $%
t<t_{in}$, when the electric field is zero. We are only interested in these
mean values for large $T$ and for $t\in Int$, when the time $t$ from the
latter interval is sufficiently large,
\begin{equation}
\sqrt{|qE|}(t-t_{in}) \gg K \gg \,\max \left\{ 1,m^{2}/|qE| \right\} \, ,
\label{time-condition}
\end{equation}
where $K$ is the cutoff introduced before in Eq.~(\ref{dec-cutoff}). Then it
is sufficient to use the $in$- and $out$-solutions in asymptotic form (\ref%
{eq:varphi-modes}) for the functions $S^{c,p}(x,x^{\prime })$ defined above.
At late times, $t>t_{out}$, the solutions $^{\pm }\psi _{\mathbf{p,}r}(x)$
reduce to free plane waves in accordance with Eq.~(\ref{free-solutions}).

\begin{figure}[h]
\includegraphics[scale=1]{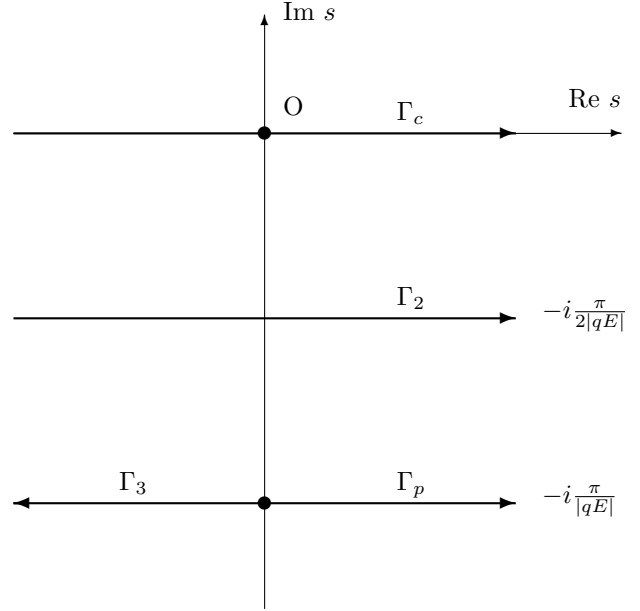}
\caption[f3]{{Contours of integration $\Gamma _{2},\Gamma
_{3},\Gamma _{c},\Gamma _{p}$}.}
\label{f3}
\end{figure}

Some components of $\langle j^{\mu }\left( t\right) \rangle ^{c,p,\bar{p}}$
and $\langle T_{\mu \nu }\left( t\right) \rangle ^{c,p,\bar{p}}$ are finite
(do not have any $T$-divergences) as $T\rightarrow \infty $. In the
expressions for these components, we can use asymptotic (as $T\rightarrow
\infty $) forms of the singular functions $S^{c,p,\bar{p}}(x,x^{\prime })$,
that is, the functions $S^{c,p,\bar{p}}(x,x^{\prime })$ in the constant
electric field. For such singular functions, the so-called Fock--Schwinger
proper time representations hold true \cite{GGG} (it should be noted that
the functions $S^{p}(x,x^{\prime })$ and $S^{\bar{p}}(x,x^{\prime })$
defined above coincide with the functions $-S^{a}(x,x^{\prime })$ and $%
-S^{p}(x,x^{\prime })$ respectively, used in the article \cite{GGG}). In the
Fock--Schwinger representations the causal propagator $S^{c}(x,x^{\prime })$
defined by Eq.~(\ref{emt1}), $S^{p}(x,x^{\prime })$ in (\ref%
{eq:pair-production-propagator}), and $S^{\bar{p}}(x,x^{\prime })$ in (\ref%
{out-out}) have the following integral representations:
\begin{align}
S^{c,p,\bar{p}}(x,x^{\prime }) = & (\gamma P+m)\Delta ^{c,p,\bar{p}%
}(x,x^{\prime }) \, ,  \notag \\
\Delta ^{c}(x,x^{\prime }) = & \int_{\Gamma_{c}} ds \, f(x,x^{\prime },s) \,
,  \notag \\
\Delta ^{p}(x,x^{\prime }) = & -\int_{\Gamma_{p}} ds \, f(x,x^{\prime},s)
\notag \\
& \quad -\Theta(x_1-x_1^{\prime }) \int_{\Gamma_{3}+\Gamma_{2} -\Gamma_{p}}
ds \, f(x,x^{\prime },s)\;,  \notag \\
\Delta^{\bar{p}}(x,x^{\prime }) = & -\int_{\Gamma_{p}} ds \,
f(x,x^{\prime},s)  \notag \\
& \quad -\Theta (x_1^{\prime }-x_1)\int_{\Gamma _{3}+\Gamma_{2}-\Gamma_{p}}
ds \, f(x,x^{\prime },s) \, ,  \label{eq:causal-proper-time}
\end{align}%
where the Fock--Schwinger kernel $f(x,x^{\prime },s)$ reads:
\begin{align}
f(x,x^{\prime },s) = &\exp \left( -i\frac{q}{2}\sigma ^{\mu \nu }F_{\mu \nu
}s\right) f^{(0)}(x,x^{\prime },s)\,,  \notag \\
f^{(0)}(x,x^{\prime },s) = & - \frac{1}{(4\pi i)^{d/2}}\frac{qEs^{-d/2+1}}{%
\sinh (qEs)}e^{iq\Lambda }e^{-im^{2}s}  \notag \\
& \, \times \exp \left[\frac{1}{4i}(x-x^\prime)qF\coth (qFs)(x-x^{\prime })%
\right] ,  \label{kernel}
\end{align}%
$\coth (qFs)$ is the matrix with the components $[\coth (qFs)]^{\mu }{}_{\nu
}$, and $\Lambda =(x_{0}+x_{0}^{\prime })(x_{1}-x_{1}^{\prime })E/2$. All
integration contours in the $s$-complex plane are shown in Fig.~\ref{f3}.
The integral along the contour $\Gamma _{c}$, that is, along the real
positive semiaxis, corresponds to the well-known Schwinger's representation
of the Feynman propagator.
%%%%%%%%%%%%%%%%%%%%%%%%%%%%%%%%%%%%%%%%%%%%%%%%%%%%%%%%%%%%%%%%%%%%%%%%
%%%%%%%%%%%%%%%%%%%%%%%%%%%%%%%%%%%%%%%%%%%%%%%%%%%%%%%%%%%%%%%%%%%%%%%

The only singular points of the kernel $f(x,x^{\prime },s)$ in the lower
half-plane outside of the origin are $s_{n}=-i\pi n/\left\vert qE\right\vert
$, $n=1,2,\ldots $.

\subsection{Peculiarities in odd dimensions\label{SS3.2}}

We shall see that in odd dimensions there exist current components related
to the so-called Chern-Simons term of the effective action. The components $%
\langle j^{1}(t)\rangle ^{p,\bar{p}}$ are ill defined in the proper time
representation due to the divergence in the limit $T\rightarrow \infty $.
They will be treated in detail at finite $T$ in Subsection \ref{SS3.5}. All
other current components $\langle j^{\mu }\left( t\right) \rangle ^{c,p,\bar{%
p}}$ are finite as $T\rightarrow \infty $ and can be determined with the
help of Eqs.~(\ref{eq:causal-proper-time}). We first consider the case of $%
d=3$ dimensions. In this case, there are two non-equivalent representations
for the $\gamma $-matrices,
\begin{equation}
\gamma ^{0}=\sigma ^{3},\;\gamma ^{1}=i\sigma ^{2},\;\gamma ^{2}=-i\left(
\pm 1\right) \sigma ^{1},  \label{eq:gamma-representations}
\end{equation}
where $\sigma ^{i}$ are the Pauli matrices and the signs\ $\pm 1$ correspond
to different fermion species, which we call as $\pm $-fermions,
respectively. By using the formula
\begin{equation*}
\exp \left( -i\frac{q}{2}\sigma ^{\mu \nu }F_{\mu \nu }s\right) =\cosh
(qEs)+\gamma ^{0}\gamma ^{1}\sinh (qEs),
\end{equation*}
one finds that
\begin{equation*}
\mathrm{tr}\left[ \gamma ^{\mu }mf(x,x,s)\right] =-i(\pm 1) 2m\sinh
(qEs)f^{(0)}(x,x,s)\delta_{\mu ,2}.
\end{equation*}
Inserting this expression and $f^{(0)}(x,x^{\prime },s)$ given by Eq.~(\ref%
{kernel}) into Eqs.~(\ref{eq:causal-proper-time}) and taking into account
that the integral along the contour $\Gamma _{3}+\Gamma _{2}-\Gamma _{p}$
does not contribute to quantities given by Eqs.~(\ref{eq:A-def}), we obtain
the following result:
\begin{align}
& \langle j^{\mu }(t)\rangle ^{c}=\pm \delta _{\mu ,2}\frac{e^{2}E}{4\pi }
\, , \qquad \langle j^{0}(t)\rangle ^{p,\bar{p}}=0 \, ,  \notag \\
& \langle j^{2}(t)\rangle ^{p,\bar{p}}=\mp \frac{e^{2}E}{4\pi ^{3/2}}\Gamma
\left( \frac{1}{2},\frac{ \pi m^{2}}{\left\vert qE\right\vert }\right) ,
\label{emt2a}
\end{align}
where $\Gamma \left( 1/2,x\right) $ is the incomplete gamma function. In $%
d\neq 3$ dimensions we find that
\begin{equation}
\langle j^{\mu }(t)\rangle ^{c}=0,\ \forall \mu ;\quad \langle j^{\mu
}(t)\rangle ^{p,\bar{p}}=0,\ \mu \neq 1.  \label{emt2b}
\end{equation}%
We see that in $d=3$ dimensions there are nonzero components $\langle j^{\mu
}(t)\rangle _{\bot }^{c,p,\bar{p}}=\delta _{\mu ,2}\langle j^{2}(t)\rangle
^{c,p,\bar{p}}$ that are orthogonal to the electric field direction. By
covariance\textbf{,} one {concludes from Eq.~(41)} that the component $%
\langle j^{\mu }(t)\rangle ^{c}$ in an arbitrary inertial reference frame is
\begin{equation*}
\langle j^{\mu }(t)\rangle ^{c}=\pm \frac{e^{2}}{8\pi }e^{\mu \alpha \beta
}F_{\alpha \beta }
\end{equation*}%
{for any electric-like field (i.e., when the magnetic field can be removed
by a Lorentz transformation). }This expression has previously been obtained
for the magnetic-like uniform electromagnetic field and related to the
additional {Chern-Simons}\ term\ in the Euler-Heisenberg effective action,
\begin{equation*}
\Gamma _{CS}=\pm \frac{e^{2}}{16\pi }\int dtd\mathbf{x}e^{\mu \alpha \beta
}F_{\alpha \beta }A_{\mu },
\end{equation*}%
in  \cite{NiemiSem83,redlich}\emph{,} while the expression for $\langle
j^{\mu }(t)\rangle _{\bot }^{p,\bar{p}}$ that can be extracted from {Eq.} ~(%
\ref{emt2a}) is new. This radiatively induced term $\Gamma _{CS}$ is the
topological mass term (Chern-Simons invariant). The Chern-Simons term $%
\Gamma _{CS}$ formally vanishes for a uniform field, but its variation with
respect to $A_{\mu }$ produces a nonvanishing current $\langle j^{\mu
}(t)\rangle ^{c}$. For the case of magnetic-like field, {i.e., when} the
electric field can be removed by a Lorentz transformation, {one has $\langle
j^{\mu }(t)\rangle =\langle j^{\mu }(t)\rangle ^{c}$ and $\langle j^{\mu
}(t)\rangle _{\bot }^{p,\bar{p}}=0$. However, in the case of an
electric-like field under consideration, we see that there is an additional
term $\langle j^{2}(t)\rangle ^{p,\bar{p}}\neq 0$ given by Eq.~~(\ref{emt2a}%
).} This is what distinguishes substantially the case of electric-like field
from the case of magnetic-like field. Thus, we see that the Chern-Simons
term is present in a properly regularized effective action for the
electric-like field with odd number of fermion species.
The signs "$\pm$" in
Eq.~(\ref{emt2a}) are opposite for each of the two possible fermion species.
The absence of similar components in expressions (\ref{emt2b}) in higher ($%
d>3$) odd dimensions is related to the fact that we consider a special case
where the magnetic field is absent. Note that in arbitrary constant field
such current components are orthogonal to the electric and magnetic field, $%
\langle j^{\mu }(t)\rangle _{\bot }^{c,p,\bar{p}}F_{\mu \nu }=0$, and
proportional to the product of all eigenvalues of the field tensor $F_{\mu
\nu }$, see the appropriate expression for $\mathrm{tr}\left[ \gamma ^{\mu
}mf(x,x,s)\right] $ in \cite{GGG}. Separating the transverse components, we
represent the vacuum current density in the final form
\begin{align}
& \langle j^{\mu }(t)\rangle =\langle j^{\mu }(t)\rangle _{\bot }+\re\langle
j^{\mu }(t)\rangle _{\Vert }^{p},\;  \notag \\
& \langle j^{\mu }(t)\rangle _{\bot }=\langle j^{\mu }(t)\rangle _{\bot
}^{c}+\langle j^{\mu }(t)\rangle _{\bot }^{p} \, ,  \label{emt2c}
\end{align}%
where finite quantities $\langle j^{\mu }(t)\rangle _{\bot }^{c,p}$ are
given by Eqs.~(\ref{emt2a}) and (\ref{emt2b}), whereas the component $%
\langle j^{\mu }(t)\rangle _{\Vert }^{p}=\delta _{\mu ,1}\langle
j^{1}(t)\rangle ^{p}$ directed along the electric field must be treated for
finite $T$. One can see from Eqs.~(\ref{mean-cr1}) and (\ref{emt2a}) that
the transverse current of created particles is absent, $j_{cr}^{2}(t)=0$,
and this conclusion holds true for arbitrary constant field in higher odd
dimensions. Then the term $\langle j^{\mu }(t)\rangle _{\bot }$ represents
transverse vacuum-polarization current, in particular for $d=3$ it has the
form
\begin{equation}
\langle j^{\mu }(t)\rangle _{\bot }=\pm \delta _{\mu ,2}\frac{e^{2}}{4\pi
^{3/2}}\gamma \left( \frac{1}{2},\frac{\pi m^{2}}{\left\vert qE\right\vert }%
\right)E ,  \label{emt2d}
\end{equation}%
where $\gamma \left( 1/2,x\right) $ is the incomplete gamma function.

The factor in front of $E$ in Eq. (\ref{emt2d}) can be considered as the
nonequilibrium Hall conductivity for large duration of electric field $%
\Delta t$ satisfying condition (\ref{time-condition}). It should be noted
that the expression  (\ref{emt2d}) is close to the one given by Eq. (7) in
\cite{DoraM11}.
The latter was obtained in the framework of one-particle WKB calculations,
using an analogy with the Landau-Zener tunneling. Note that it is quite
difficult in the framework of the one-particle theory to distinguish between
the current of real particles, which remains after switching off of the
electric field, and the pure vacuum-polarization current, which disappears
in this case, although there is no dissipation in the model. We stress that
the latter current vanishes together with the electric field. Therefore, in
this problem, a use of the analogy with the tunneling, i.e., with
Schwinger's pair-production rate, should be justified.
In fact, our exact result (\ref{emt2d}), obtained in the
framework of QED, is such a justification of the one-particle calculation
\cite{DoraM11}, at the same time it shows the limits of its validity,
namely, it holds true only before the switching off the electric field. The
quantity $\langle j^{\mu }(t)\rangle _{\bot }^{p,\bar{p}}$ is exponentially
small for weak electric fields, $m^{2}/\left\vert qE\right\vert \gg 1$, in
which case $\langle j^{2}(t)\rangle \approx \langle j^{2}(t)\rangle ^{c}$.
On the other hand, in the strong-field limit, $m^{2}/\left\vert
qE\right\vert \ll 1$, both contributions are comparable, $\langle
j^{2}(t)\rangle ^{p}\approx -\langle j^{2}(t)\rangle ^{c}$, and then
\begin{equation}
\langle j^{2}(t)\rangle \approx \pm \frac{e^{3/2}Em}{2\pi \sqrt{\left\vert
E\right\vert }}\,.  \label{emt3d}
\end{equation}%
The conductivity that follows from eq. (\ref{emt3d}) coincides with the one
given by Eq. (9) in  \cite{DoraM11}.

As it is known, the term $\langle j^{2}(t)\rangle ^{c}$ is related to the
standard effective action and therefore to probability amplitudes of
processes, while the term $\langle j^{2}(t)\rangle $ presents a contribution
to mean values which, in general, are quite different from such amplitudes.
We see that $\langle j^{2}(t)\rangle ^{c}$ and the probability amplitudes
remain unchanged as $m\rightarrow 0$. In contrast to the behavior of
quantity $\langle j^{2}(t)\rangle ^{c}$, the mean value $\langle j^{2}(t)\rangle
$ tends to $0$ as $m\rightarrow 0$. It is important to note that these
constant values of both $\langle j^{\mu }(t)\rangle _{\bot }^{c}$ and $%
\langle j^{\mu }(t)\rangle _{\bot }^{p}$ are obtained in the limit $%
T\rightarrow \infty $, that is, Eqs.~(\ref{emt2a}) hold true for instants of
time $t$ before the electric field is switched off. At early, $t<t_{in}$,
and late times, $t>t_{out}$, the $T$-constant electric field is absent, and
one can see from the exact formula (\ref{eq:A-def}) that the vacuum
polarization current vanishes in this case, $\langle j_{\mu }(t)\rangle
^{c}=\langle j^{\mu }(t)\rangle _{\bot }^{p}=0$. These results can be
generalized to the case of massless fermions in an electric-like constant
electromagnetic field in higher odd dimensions, when the electric field
cannot be removed by the Lorentz transformation and all eigenvalues of the
field tensor $F_{\mu \nu }$ are different from zero. In this case when $%
t<t_{out}$, the transverse component of vacuum polarization current $\langle
j^{\mu }(t)\rangle _{\bot }^{c}$ is {nonzero}, however, the total transverse
mean value $\langle j^{\mu }(t)\rangle _{\bot }$ is equal to zero. Thus, the
vacuum polarization in the electric field in odd dimensions is qualitatively
different for mean values and amplitudes of processes. This is what
distinguishes substantially the case of electric-like field from the case of
magnetic-like field.

\subsection{EMT\label{SS3.3}}

Using representations (\ref{eq:A-def}) and (\ref{eq:causal-proper-time}), we
obtain components of the EMT as follows:
\begin{align}
& \langle T_{\mu \nu }(t)\rangle ^{c,p,\bar{p}}=0 \, , \quad \mu \neq \nu \,
,  \label{emt3} \\
& \re \langle T_{00}(t)\rangle ^{c}=-\re \langle T_{11}(t)\rangle ^{c}=E%
\frac{\partial \re \mathcal{L}(t)}{\partial E}-\re \mathcal{L}(t)\,,  \notag
\\
& \re \langle T_{ii}(t)\rangle ^{c}=\re \mathcal{L}(t) \, , \quad
i=2,3,\dots ,D,  \label{emt4}
\end{align}
where
\begin{align}
& \mathcal{L} = \frac{1}{2}\int_{0}^{\infty }\frac{ds}{s}\mathrm{tr}f(x,x,s)
\, ,  \notag \\
& \mathrm{tr}f(x,x,s)=2^{[d/2]}\cosh (qEs)f^{(0)}(x,x,s).  \label{ELa}
\end{align}
It should be noted that the diagonal elements $\langle T_{\mu \mu
}(t)\rangle ^{p}$ are ill defined in the proper time representation due to
the divergence at the limit $T\rightarrow \infty $ and will be treated at
finite $T$ in Subsec.  \ref{SS3.5}. The quantity $\mathcal{L}$ in (\ref{ELa}) can be
identified as the non-renormalized one-loop effective Euler-Heisenberg
Lagrangian of the Dirac field in an uniform electric field. It is the
density of the one-loop effective action $W,$%
\begin{equation}
W=\int \mathcal{L}dtd\mathbf{x}=-i\ln c_{v}.  \label{EA}
\end{equation}%
that is defined in general via the vacuum-to-vacuum amplitude (\ref%
{eq:causal-propagator}). Its imaginary part represents the vacuum-to-vacuum
probability, as follows:
\begin{equation}
\left\vert c_{v}\right\vert ^{2}=e^{-2\im W}.  \label{Pv}
\end{equation}
It is a global physical quantity. It is free of ultraviolet divergences,
because creation of pairs with infinitely large momenta is suppressed. The
Bogolyubov transformation between $in$- and $out$- creation and annihilation
operators allows one to relate $c_{v}$ with the $g$ coefficients from (\ref%
{eq:in-out-g}) \cite{FGS,GavGT06}:
\begin{equation*}
c_{v}=\exp \left\{ \mathrm{tr}\ln g(_{-}\mid ^{-})^{\ast }\right\} \,.
\end{equation*}

For the $T$-constant field, when $T$ satisfies the stabilization conditions (%
\ref{stabilization-condition}), the probability (\ref{Pv}) can be expressed
via total number of particles created (\ref{n-cr}) in the following form:
\begin{equation}
\ln |c_v|^2=-\rho \aleph \, , \quad \rho =\sum_{n=0}^{\infty
}(n+1)^{-d/2}\exp \left( -n\pi m^{2}/|qE| \right) ,  \label{ImW}
\end{equation}%
see \cite{GG95}. It follows from (\ref{ImW}) that $\aleph \sim TV_{(d-1)}$,
then $\ln \left\vert c_{v}\right\vert ^{2}$ is proportional to the spatial
volume $V_{(d-1)}$ and the field duration $T$. It can be verified that the
result (\ref{ImW}) can be obtained with the help of $\im \mathcal{L}$ (\ref%
{ELa}), provided that the integral over the time $t$ is identified with the
field duration $T$, $\int dt=T,$ \cite{GGG}. Thus, $\im W$ is finite for
finite values of $V_{(d-1)}$ and $T$. Note that in the strong-field case, $%
m^{2}/\left\vert qE\right\vert \ll 1$, for $d\geq 3$, the leading term in
the r.h.s. of (\ref{ImW}) is the ordinary Riemann zeta function, $\rho
=\zeta _{R}\left( d/2\right)$. For $d=2$, the leading term is $\rho =\ln
\left( \left\vert qE\right\vert /m^{2}\right) $. That means that in $d=2$
dimensions the vacuum-to-vacuum probability of massless fermions in
quasiconstant electric field is ill defined. In this case, the mass term $%
m\neq 0$ has to be considered as an infrared cutoff.

Both the vacuum mean values of the total current density and of the EMT are
represented as sums of two contributions: $\re\langle j^{\mu }(t)\rangle ^{c}
$ (resp. $\re\langle T_{\mu \nu }(t)\rangle ^{c}$ and $\re\mathcal{L}(t)$)
that can be associated with the vacuum polarization, while $\langle j^{\mu
}(t)\rangle ^{p}=$ $j_{cr}^{\mu }(t)+\langle j^{\mu }(t)\rangle ^{\bar{p}}$
(resp. $\langle T_{\mu \nu }(t)\rangle ^{p}=T_{\mu \nu }^{cr}(t)+\langle
T_{\mu \nu }(t)\rangle ^{\bar{p}}$) can be associated with the pair creation
due to the term $j_{cr}^{\mu }(t)$ (resp. $T_{\mu \nu }^{cr}(t)$). The
latter follows from the fact that $S^{p}(x,x^{\prime })$ with all its
derivatives, given by Eq.~(\ref{eq:pair-production-propagator}), is
exponentially small due to the smallness of $|g(_{+}\mid ^{-})|$, when the
electric field is weak, $m^{2}/\left\vert qE\right\vert /\gg 1$. Of course,
in general, such unambiguous division of physical quantities due to particle
creation and the vacuum polarization is not possible. However, it can be
done in some specific cases and for some specific quantities. It is clear
that $j_{cr}^{\mu }(t)$ and $T_{\mu \nu }^{cr}(t)$ depend on the history of
the process and retain their latest values at $t>t_{out}$ . On the other
hand, we see from Eqs.~(\ref{emt2a}) and (\ref{emt4}) that the nonzero real
parts of the quantities $\langle j^{\mu }(t)\rangle ^{c},$ $\re\langle
T_{\mu \nu }(t)\rangle ^{c}$, and $\langle j^{\mu }(t)\rangle _{\bot }^{p}$
are time-independent for the time $t\in Int$. At early $t<t_{in}$ and late
times $t>t_{out}$ , we have $\langle j^{\mu }(t)\rangle ^{c}=\langle j^{\mu
}(t)\rangle _{\bot }^{p}=0$ and the mean $\langle T_{\mu \nu }(t)\rangle ^{c}
$ is reduced to its free value for $E=0$. Thus, we see that  $\langle j^{\mu
}(t)\rangle ^{c}$, $\langle j^{\mu }(t)\rangle _{\bot }^{p}$, and $\re%
\langle T_{\mu \nu }(t)\rangle ^{c}$ depend on the electric field at the
time $t$, but do not depend on the history of the process, that is, they are
local quantities and represent the vacuum polarization contribution. We
continue to study each of these local and nonlocal terms independently in
the following sections.

\subsection{Mean values: renormalization\label{SS3.4}}

The integrals (\ref{emt4}) are divergent due to the real part of the
effective Lagrangian (\ref{ELa}) which is ill defined. This real part must
be regularized and renormalized. In low dimensions, $d\leq 4$, $\re \mathcal{%
L}$ can be regularized in the proper-time representation and renormalized by
the Schwinger renormalizations of the charge and the electromagnetic field
\cite{schwinger}. In higher dimensions, a different approach is required. 
Note that in the case of $d>4$\ dimensions, plane QED is rather unrealistic
system, however, it is common to use it as a simple (Abelian) model to
consider the qualitative behavior of a quantum\ gauge field theory as a
function of $d$, see, for example, \cite{NiemiSem83,redlich}. One can 
treat it as an effective theory given by one-loop effective action that can
be regularized and renormalized in some appropriate way.
Of course, exact meaning of finite and divergent terms of
effective action at $d>4$ can be understood only from the corresponding
fundamental theory. In our article we consider the strong-field asymptotic
behavior of the one-loop effective action. One can see that this asymptotic
behavior is insensitive to methods of regularization and renormalization,
see, for example,  \cite{dunne,wipf}. At $d>4$  one can give a precise
meaning and calculate the one-loop effective action using zeta-function
regularization \cite{elizalde}.
An application of
this method to the case of a uniform magnetic field and self-dual field in
arbitrary dimensions is described in detail in \cite{wipf}. It can be shown
that for $d\leq 4$ such a renormalization is equivalent to the above
mentioned Schwinger's renormalization, see \cite{wipf}. Let us consider the
application of this technique for the case of a constant uniform electric
field in arbitrary dimensions of interest here.

First, we remind that the effective action $W$ from (\ref{EA}) can be
represented as the functional determinant,%
\begin{equation}
W=-\frac{i}{2}\ln \det M^{2} \, ,\quad M^{2}=m^{2}-i0-P^{2}+\frac{q}{2}%
\sigma ^{\mu \nu }F_{\mu \nu }.  \label{r1}
\end{equation}%
The operator $M^{2}$ becomes elliptic when it is continued to the Euclidean
space, $M^{2}\rightarrow \tilde{M}^{2}$, by means of the replacements $%
t\rightarrow -i\eta $,\ $\partial _{0}\rightarrow i\partial _{\eta }$, and\ $%
qE\rightarrow iB.$ Then the functional determinant is well-defined with the
help of the zeta-function $\zeta ^{\left( d\right) }\left( s\right) $ in $d$
dimensions,
\begin{align}
& \ln \det \tilde{M}^{2}=-\left. \frac{d\zeta^{(d)}(s)}{ds}\right\vert
_{s=0} \, ,  \notag \\
& \zeta^{(d)}(s) = \Gamma^{-1}(s) \int_{0}^{\infty }du \, u^{s-1}K\left(
u\right) \, ,  \label{r2}
\end{align}
where $K\left( u\right) $ is the heat kernel,
\begin{align*}
& K\left( u\right) =\int d\eta \, d\mathbf{x} \, \mathrm{tr}f_{Eucl}(x,x,u)
\, , \\
& f_{Eucl}(x,x,u)=\left\langle \eta ,\mathbf{x}\left\vert \exp \left( -u%
\tilde{M}^{2}/\mu ^{2}\right) \right\vert \eta ,\mathbf{x}\right\rangle .
\end{align*}%
Here the quantity $\mu $ is a renormalization scale, which is introduced to
keep the zeta-function dimensionless. The dependence on $\mu $ of the
functional determinant corresponds to a finite renormalization.

The effective action $W$ can be written as the following integral over the
Euclidean space time volume: $W=-i\int \mathcal{L}d\eta d\mathbf{x}.$ It is
real when it is continued to the imaginary electric field $qE\rightarrow iB$%
, so that $\tilde{W}=\left. W\right\vert _{qE=iB}$, $\im \tilde{W}=0$. Note
that for $d\geq 3$, $\tilde{W}$ coincides with the effective action of the
magnetic field $B/q$. Consequently, we obtain a regularized and finite form
of the effective Lagrangian of a constant electric field in an arbitrary $d$%
-dimensions as follows:
\begin{equation}
\re \mathcal{L}_{reg}=\re \left. \mathcal{\tilde{L}}\right\vert _{B=-iqE} \,
, \quad \mathcal{\tilde{L}}=-\frac{1}{2}\left. \frac{d\zeta^{(d)}(s) } {ds}%
\right\vert _{s=0}\Omega _{\left( d\right)}^{-1} \, ,  \label{r5}
\end{equation}%
where $\Omega _{(d)}=\int d\eta \, d\mathbf{x}$ is the Euclidean space-time
volume in $d$ dimensions.

An explicit form of the quantity $\mathrm{tr}f_{Eucl}(x,x,u)$ can be
extracted from the quantity $\mathrm{tr}f(x,x,s)$ (given by (\ref{kernel})
and (\ref{ELa})) as follows:
\begin{equation}
\mathrm{tr}f_{Eucl}(x,x,u)=-\left. \mathrm{tr}f\left( x,x,-\frac{iu}{\mu^{2}}
\right)\right\vert _{qE=iB} \, .  \label{r6}
\end{equation}
Then we have:
\begin{align}
\zeta^{(2)}\left( s\right) & = \frac{\Omega _{\left( 2\right) }}{2\pi }
\frac{1}{\Gamma(s)} \int_{0}^{\infty }u^{s-1}B\coth \left( \frac{Bu}{\mu^2}
\right) \text{e}^{-m^{2}u/\mu^2} du \, ,  \notag \\
\zeta^{(d)}(s) & = \frac{\Omega _{\left( d\right) }2^{[d/2]-1}}{\Omega
_{\left( 2\right) }}\left( \frac{\mu ^{2}}{4\pi }\right) ^{d/2-1}\frac{%
\Gamma \left( s-d/2+1\right) }{\Gamma (s) }  \notag \\
& \qquad \times \zeta^{(2) }\left( s-d/2+1\right) \, , \quad \text{for } d>2
\, .  \label{r7}
\end{align}

If $B>0$, the function $\zeta^{(2)}(s)$ can be written in terms of Hurwitz
zeta-function as follows \cite{elizalde,wipf}:
\begin{equation}
\zeta^{(2)}(s) =\left\{
\begin{array}{l}
\frac{\Omega_{(2)}B}{2\pi}\left[ 2\left( \frac{2B}{\mu ^{2}}\right)^{-s}
\zeta_H \left( s,1+\frac{m^{2}}{2B}\right) + \left( \frac{m^{2}}{\mu ^{2}}%
\right) ^{-s}\right] , \\
\hfill \text{for } m \neq 0 \, ; \\
\Omega_{(2)}\frac{B}{\pi }\left( \frac{2B}{\mu ^{2}}\right)^{-s}\zeta_{R}(s)
\, , \quad \text{for }m=0 \, .%
\end{array}%
\right.  \label{r9}
\end{equation}

The Hurwitz zeta-function is defined as an analytic continuation of the
series%
\begin{equation*}
\zeta_H(s;x) =\sum_{n=0}^{\infty }(n+x)^{-s} \, , \quad\re s>1 \, ,
\end{equation*}
to the entire complex plane of $s$. One can see that $\zeta _{R}(s) =\zeta
_{H}(s,1)$ is the ordinary Riemann zeta-function. For odd $d$, and taking
into account that $\Gamma ^{-1}(s) \approx s$ at $s\approx 0$, we get from (%
\ref{r7}) that%
\begin{equation}
\left. \frac{d\zeta ^{\left( d\right) }\left( s\right) }{ds}\right\vert
_{s=0}=\left. \Gamma \left( s\right) \zeta ^{\left( d\right) }\left(
s\right) \right\vert_{s=0} \, .  \label{r10}
\end{equation}%
By using Eq.~(\ref{r5}) we find a finite expression for $\re \mathcal{L}%
_{reg}$ in arbitrary dimensions. Then we have to implement an additional
finite renormalization of the cosmological constant and the electric charge
\cite{wipf}. For $d>4$, a finite renormalization of some high-dimensional
quantities could be needed. Thus, we obtain the final form of the
renormalized effective Lagrangian $\re \mathcal{L}_{ren}$. The corresponding
final forms for $\mathcal{\tilde{L}}$ and for the $B$ field in $d=2,3,4$
dimensions are treated in detail in \cite{wipf}.

We are interested in the case of a very strong field, $m^{2}/\left\vert
qE\right\vert \ll 1$. In this case the leading contribution to $\re\mathcal{L%
}_{ren}$ is given by $\re \mathcal{L}_{reg}$ in (\ref{r5}), where $%
m\rightarrow 0$. Using (\ref{r7}), (\ref{r9}), and (\ref{r10}), we find that
this contribution has the form%
\begin{equation}
\re \mathcal{L}_{ren}\approx \frac{1}{2}\re \left\{
\begin{array}{l}
\left[ \ln \left( \frac{B}{\mu ^{2}}\right) \frac{\zeta^{(d)}(0)}{%
\Omega_{(d)}} \right]_{B=-iqE} \, , \text{ for even } d, \\
- \left[ \left. \Gamma(s) \frac{\zeta^{(d)}(s)}{\Omega_{(d)}} \right\vert
_{s=0}\right] _{B=-iqE} , \text{ for odd } d.%
\end{array}%
\right.  \label{r11}
\end{equation}%
In particular,
\begin{widetext}
\begin{align}
\re \mathcal{L}_{ren} \approx \left\{
\begin{array}{llll}
&-\re \left. \left[ \frac{B}{4\pi }\ln \left( \frac{B}{\mu ^{2}}\right) %
\right] \right\vert _{B=-iqE}  = \frac{\left\vert qE\right\vert }{4} \, , & \text{for }d=2 \, , \\
& -\re \left. \left[ \frac{1}{2\pi ^{2}}\left( \frac{B}{2}\right)
^{3/2}\zeta _{R}\left( \frac{3}{2}\right) \right] \right\vert _{B=-iqE}  =
\frac{\left\vert qE\right\vert ^{3/2}}{8\pi ^{2}}\zeta _{R}\left( \frac{3}{2}%
\right) \, , & \text{for }d=3 \, , \\
& \re \left. \left[ \frac{B^{2}}{24\pi ^{2}}\ln \left( \frac{B}{\mu ^{2}}%
\right) \right] \right\vert _{B=-iqE}  =-\frac{\left( qE\right) ^{2}}{24\pi
^{2}}\ln \left( \frac{\left\vert qE\right\vert }{\mu ^{2}}\right) \, , & \text{for }d=4 \, .
\end{array}%
\right.   \label{r12}
\end{align}
\end{widetext}
For $d=4$, the obtained $\re \mathcal{L}_{ren}$ coincides with the already
known result \cite{dunne}. For the magnetic field, the leading term in the
form of the second Eq.~(\ref{r12}) has previously been obtained in \cite%
{redlich} for $d=3$. Note that despite the fact that the nonzero vacuum
current $\langle j^{\mu }(t)\rangle ^{c}$ is present in $d=3$ dimensions,
the Chern-Simons term vanishes for the constant field strength \cite{redlich}%
. In general, we have for a very strong electric field that
\begin{equation*}
\re \mathcal{L}_{ren}\sim \left\{
\begin{array}{ll}
|qE|^{d/2} \, , & d\neq 4n \, , \\
\left\vert qE\right\vert ^{d/2}\ln \left( |qE|/\mu ^{2}\right) \, , & d=4n
\, .%
\end{array}%
\right.
\end{equation*}%
In contrast to the electric field case, where the logarithmic factor $\ln
\left( |qE|/\mu ^{2}\right) $ appears only for $d=4n$, in a strong magnetic
field it is present in any even dimension: $\re \mathcal{L}_{ren}\sim
B^{d/2}\ln (B/\mu ^{2})$ for even $d$, and $\re \mathcal{L}_{ren}\sim B^{d/2}
$ for odd $d$. In the framework of the on-shell renormalization of massive
theory, we have to set $\mu =m$.

Thus, we have obtained the renormalized mean values of EMT components in the
following form:
\begin{align}
\re \, \langle T_{00}(t)\rangle _{ren}^{c} & =-\re \langle T_{11}(t)\rangle
_{ren}^{c}  \notag \\
& =E\frac{\partial \re \mathcal{L}_{ren}(t)}{\partial E}-\re \mathcal{L}%
_{ren}(t) \, ,  \notag \\
\re \, \langle T_{ii}(t)\rangle _{ren}^{c} & =\re \mathcal{L}_{ren}(t) \, ,
\quad \ i=2,3,\dots ,D,  \label{L-ren}
\end{align}%
where $\mathcal{L}\left( t\right) _{ren}$ at $t\in Int$ is given by (\ref%
{r11}). Thus, in the strong-field case, the quantities (\ref{L-ren}) have
the following behavior:
\begin{equation}
\re \, \langle T_{\mu \mu }(t)\rangle _{ren}^{c}\sim \left\{
\begin{array}{l}
\left\vert qE\right\vert ^{d/2}\ln \left( \left\vert qE\right\vert /\mu
^{2}\right) \, , \quad d=4n \, , \\
\left\vert qE\right\vert^{d/2} \, , \quad d\neq 4n \, ,\ n \in \mathbb{N} \,
.%
\end{array}
\right.  \label{emt-lead}
\end{equation}%
At early $t<t_{in}$ and late $t>t_{out}$ times, we have $\re\mathcal{L}%
(t)_{ren}=0$ and $\re \langle T_{\mu \nu }(t)\rangle _{ren}^{c}=0$. Taking
into account (\ref{L-ren}), we find the final form for the vacuum mean
values of the EMT
\begin{equation}
\langle T_{\mu \nu }(t)\rangle _{ren}= \re \, \langle T_{\mu \nu }(t)\rangle
_{ren}^{c}+\re \, \langle T_{\mu \nu }(t)\rangle ^{p},  \label{emt-ren}
\end{equation}%
where, according to (\ref{emt3}), the off-diagonal elements are equal to
zero, {and the diagonal elements of $\re \, \langle T_{\mu \nu }(t)\rangle
_{ren}^{c}$ are given by \eqref{L-ren}; the diagonal components of $\re \,
\langle T_{\mu \nu }(t)\rangle ^{p}$ are studied in detail in the next
section.}

\subsection{Mean values: pair-creation contributions\label{SS3.5}}

We know that the mean values $\langle j^{1}(t)\rangle ^{p,\bar{p}}$ and $%
\langle T_{\mu \mu }(t)\rangle ^{p,\bar{p}}$ are divergent as $T\rightarrow
\infty$. Therefore, in the representation (\ref{eq:A-def}), they have to be
considered always at finite $T$. Let us evaluate these quantities in the
large $\tau $-limit approximation with $\tau =\sqrt{\left\vert qE\right\vert
}(t-t_{in})$. Consider the time $t\in Int$ for which the interval $t-t_{in}$
satisfies condition (\ref{time-condition}) and the cutoff approximations%
\begin{align}
& |\mathbf{p}_{\perp }|\leq \sqrt{\left\vert qE\right\vert }\left[ \sqrt{%
\left\vert qE\right\vert }\left( t-t_{in}\right) -K\right] ^{1/2}\,,  \notag
\\
&t_{in}+K/\sqrt{\left\vert qE\right\vert }\leq p_{1}/qE\leq t-K/\sqrt{%
\left\vert qE\right\vert }\,,  \label{domain-def}
\end{align}%
hold true. In Eqs.~(\ref{domain-def}) it is taken into account that physical
observables in the time moment $t\in Int$ are affected by the electric field
that acted for the time $[t_{in},t]$. Let us call the corresponding region
in the momentum space by $\Omega (t)$. Using the transformation %
\eqref{eq:in-out-g}, one can represent the propagator $S^{p}$ in Eq.~%
\eqref{eq:pair-production-propagator} in terms of $out$-solutions. In the
asymptotic regime, as $z\rightarrow \infty $, the Weber functions have the
following asymptotic expansion:
\begin{equation*}
D_{\nu }(z)\simeq z^{\nu }\exp \left( -z^{2}/4\right) \left[ \sum_{n=0}^{N}%
\frac{\left( -\frac{\nu }{2}\right) _{n}\left( \frac{1}{2}-\frac{\nu }{2}%
\right) _{n}}{n!(-z^{2}/2)^{n}}\right] ,
\end{equation*}%
which is valid for $|\arg z|<3\pi /4$. Then, keeping only the zeroth ($n=0$)
term in last equation, we obtain for $x\simeq x^{\prime }$:
\begin{align}
S^{p}(x,x^{\prime }) = & (\gamma P+M)\Delta ^{p}(x,x^{\prime }) \, , \\
\Delta^{p}(x,x^{\prime })= & -i\int d\mathbf{p}\,2\left\vert
qEt-p_{1}\right\vert \aleph _{\mathbf{p}}e^{i\mathbf{p}\cdot (\mathbf{x}-%
\mathbf{x} ^{\prime })}  \notag \\
& \quad \times \left[ ^{+}\varphi _{\mathbf{p,}-1}(t){}^{+}\varphi _{\mathbf{%
p,}-1}^{\ast }\left( t^{\prime }\right) \right.  \notag \\
& \left. \qquad +^{-}\varphi _{\mathbf{p,}+1}(t){}^{-}\varphi _{\mathbf{p,}%
+1}^{\ast }\left( t^{\prime }\right) \right] \,.  \label{Delta-def}
\end{align}%
Considering the large $\tau $-limit in representations (\ref{eq:A-def}), the
domain of integration in (\ref{Delta-def}) can be restricted to the region $%
\Omega (t)$ described by the inequalities (\ref{domain-def}). Using Eq.~%
\eqref{eq:Np-asymptotic} for the differential mean values $\aleph _{\mathbf{p%
}}$, we obtain:
\begin{align}
\Delta ^{p}(x,x^{\prime })= & -i\int_{t_{in}+K/\sqrt{\left\vert
qE\right\vert }}^{t-K/\sqrt{\left\vert qE\right\vert }}h_{\perp }(\mathbf{x}%
_{\perp },\mathbf{x}_{\perp }^{\prime })h_{\parallel }(x_{\parallel
},x_{\parallel }^{\prime })d\tilde{t}\,,  \notag \\
h_{\parallel }(x_{\parallel },x_{\parallel }^{\prime }) = & \frac{1}{t-%
\tilde{t}}\,\text{e}^{ip_1\left( x_{1}-x_{1}^{\prime }\right)} \cos\left\{
\frac{1}{2}\left[ \xi (t^{\prime })^{2}-\xi (t)^{2}\right] \right\} ,  \notag
\\
h_{\perp }(x_{\perp },x_{\perp }^{\prime }) = & \frac{|qE|^{d/2-1}}{(2\pi
)^{d-1}}  \notag \\
& \times \exp \left( -\frac{\pi m^{2}}{|qE|}-\frac{(\mathbf{x}_{\perp }-%
\mathbf{x}_{\perp }^{\prime})^2 |qE|}{4\pi}\right) \, ,  \label{Delta-eff}
\end{align}
where $p_{1}=qE\tilde{t}$.

It follows from Eqs.~(\ref{eq:A-def}) and (\ref{Delta-eff}) that
\begin{align}
& \langle j^{1}(t)\rangle ^{p}=-iq2^{[d/2]}\left. P_{1}\Delta
^{p}(x,x^{\prime })\right\vert _{x=x^{\prime }}\,,  \notag \\
& \langle T_{\mu \mu }(t)\rangle ^{p}=i2^{[d/2]}\left. P_{\mu }^{2}\Delta
^{p}(x,x^{\prime })\right\vert _{x=x^{\prime }}\,.  \label{j-GF}
\end{align}%
Integrating over $p_{1}$, we obtain the following result in the large $\tau $%
-limit (with $\tau =\sqrt{\left\vert qE\right\vert }\left( t-t_{in}\right) )$%
:
\begin{align}
\langle j^{1}(t)\rangle^{p} & = 2e\;\mathrm{sgn}\left( E\right) r^{cr}\left[
t-t_{in}+|qE|^{-1/2}O\left( K\right) \right] ,  \label{p-current} \\
\langle T_{00}(t) \rangle ^{p} & = \langle T_{11}(t) \rangle ^{p}  \notag \\
& =\left\vert qE\right\vert r^{cr}\left[ t-t_{in}+\left\vert qE\right\vert
^{-1/2}O\left( K\right) \right] ^{2},  \notag \\
\langle T_{ii}(t)\rangle ^{p} & = \pi ^{-1}r^{cr}\left\{ \ln \left[ \sqrt{%
\left\vert qE\right\vert }\left( t-t_{in}\right) \right] +O\left( \ln
K\right) \right\} ,  \notag \\
& \qquad \text{for }i=2,3,\dots ,D,  \label{p-emt}
\end{align}%
where $r^{cr}$ is given by (\ref{n-cr}). We see that all the leading
contributions given by (\ref{p-current}) and (\ref{p-emt}) are real. The
quantities $\langle j^{1}(t)\rangle ^{p}$ and $\langle T_{\mu \mu
}(t)\rangle ^{p}$ depend on the time interval $\left( t-t_{in}\right) $ of
the electric field action, showing that they are global quantities.

Now we estimate, for $t\approx t_{out}$, the current density and EMT of
created particles, $j_{cr}^{\mu }(t)$ , $T_{\mu \nu }^{cr}(t)$, given by
Eqs.~(\ref{mean-cr1}). At this time instant, the solutions $^{\pm }\psi _{%
\mathbf{p,}r}(x)$ reduce to free particle plane waves in agreement with Eq.~(%
\ref{free-solutions}). Thus, taking into account representation (\ref%
{out-out}), one can see that the quantities $\langle j^{1}(t)\rangle ^{\bar{p%
}}$ and $\langle T_{\mu \mu }(t)\rangle ^{\bar{p}}$ from (\ref{eq:A-def})
can be neglected in the large $\tau $-limit. Using Eqs.~(\ref{p-current})
and (\ref{p-emt}), we obtain:
\begin{align}
& j_{cr}^{\mu}(t) \simeq \delta _{\mu ,1}\langle j^{1}(t) \rangle ^{p} \, ,
\notag \\
& T_{\mu \nu }^{cr}(t) \simeq \delta_{\mu ,\nu }\langle T_{\mu \mu }\left(
t\right) \rangle^{p} \, , \quad \text{at } t \approx t_{out} \, ,
\label{mean-cr2}
\end{align}
where Eqs.~(\ref{emt2a}), (\ref{emt2b}), and (\ref{emt3}) were taken into
account. For $t>t_{out}$ (after the electric field is switched off), the
quantities $j_{cr}^{\mu }(t)$ and $T_{\mu \nu }^{cr}(t)$ are constant and
retain their values at $t_{out}$; for this reason, we always have to set $%
t-t_{in}$ $=T$ in such cases. The renormalized vacuum polarization
contributions in expressions (\ref{emt2c}) and (\ref{emt-ren}) vanish in the
absence of external field. Therefore, for $t>t_{out}$, the vacuum mean
values $\langle j^{\mu }(t)\rangle$ and $T_{\mu \nu }(t)\rangle _{ren}$
represent the mean current density and EMT of pairs created by the complete $%
T$-constant electric field:
\begin{equation}
\langle j^{\mu }(t)\rangle =j_{cr}^{\mu }(t_{out}), \quad \langle T_{\mu \nu
}(t)\rangle _{ren}=T_{\mu \nu }^{cr}(t_{out}), \quad t>t_{out}.
\label{mean-cr3}
\end{equation}

We see that the current density of created particles $j_{cr}^{\mu }(t)$ is
directed along the direction of the electric field. In $d=3$ dimensions, in
contrast to this almost obvious property, for $t_{in}$ $<t<t_{out}$, the
mean current density of each massive fermion specie deviates from the
direction of the electric field. Indeed,
\begin{equation}
\frac{\langle j^{2}\left( t\right) \rangle }{\langle j^{1}\left( t\right)
\rangle }=\pm \frac{\sqrt{\pi }}{2}\gamma \left( \frac{1}{2},\frac{\pi m^{2}%
}{\left\vert qE\right\vert }\right) \frac{\exp \left(\pi m^{2}/\left\vert
qE\right\vert \right)}{\sqrt{\left\vert qE\right\vert }\left( t-t_{in}\right)%
} \, ,  \label{angle}
\end{equation}%
where $\left( \sqrt{\left\vert qE\right\vert }T\right) ^{-1}\ll 1$ according
to stabilization condition (\ref{stabilization-condition}). One finds a
similar deviation of the mean current density of each massive fermion
species in higher odd dimensions for $t_{in}$ $<t<t_{out}$ in the case of an
electric-like constant electromagnetic field, when all eigenvalues of the
field tensor $F_{\mu \nu }$ are different from zero. The total mean current
density of an even number of fermion species is directed along the direction
of the electric field, since the contributions of the $\pm$-fermions differ
only in sign.

One can see from Eq.~(\ref{p-current}) that $r^{cr}T$ is the total number
density of pairs created and accelerated during the time $T$ to velocities
nearly the speed of light. It coincides with the quantity $n^{cr}$ obtained
in a different manner in (\ref{n-cr}). The quantity $T_{00}^{cr}=\left\vert
qE\right\vert Tn^{cr}$ is the mean energy density of pairs created at any
time instant $t\in Int$ with zero longitudinal kinetic momentum and then
uniformly accelerated to kinetic momenta from zero to the maximum $%
\left\vert qE\right\vert T$, so that $\left\vert qE\right\vert T/2$ is the
mean kinetic momentum per particle. The energy density $T_{00}^{cr}$ is
equal to the pressure $T_{11}^{cr}$ along the direction of the electric
field. This equality is a natural equation of state for noninteracting
particles accelerated by an electric field to relativistic velocities. The
momentum density of created pairs $T_{0i}^{cr}$ is zero due to the symmetry
between particle and antiparticle distributions.

The vacuum mean values (\ref{emt2c}) and (\ref{emt-ren}) for $t\in Int$ are
sources in equations of motion for mean electromagnetic and metric fields
respectively. It should be noted that only when the time $t$ is sufficiently
close to $t_{out}$, $\left( t_{out}-t\right) /T\ll 1$, the differences
between the densities $\langle j^{1}(t)\rangle ^{p}$ and $\langle T_{\mu \nu
}(t)\rangle ^{p}$ and the respective densities $j_{cr}^{1}\left( t\right) $
and $T_{\mu \nu }^{cr}\left( t\right) $ of final pairs created, can be
neglected and the interpretation of particles at $t$ as final $out$%
-particles is correct.

In the general case when the time $t$ is not close to $t_{out}$, there is an
essential difference between the definition of the vacuum at $t<t_{out}$ and
the final vacuum state $|0,out\rangle $ at $t_{out}$. That is why the
quantities $\langle j^{1}(t)\rangle ^{p}$ and $\langle T_{\mu \nu
}(t)\rangle ^{p}$ have nothing to do with characteristics of final $out$%
-particles. They present contributions to mean values due to vacuum
instability which depend on the history of the process, that is, they are
global quantities, in contrast of the local quantities $\langle j^{\mu
}(t)\rangle ^{c}$, $\langle j^{\mu }(t)\rangle _{\bot }^{p}$, and $\re %
\,\langle T_{\mu \nu }(t)\rangle ^{c}$. In the general case, $r^{cr}$ in
Eqs.~(\ref{p-current}) and (\ref{p-emt}) is the total number density of
excited states per unit of time. For example, the longitudinal component of
the mean current density $\langle j^{1}(t)\rangle ^{p}$ increases linearly
as ($t-t_{in}$ $)$ grows, since the decoherence does not take place for $%
t<t_{out}$. We note that $\langle j^{1}(t_{out}$ $)\rangle \approx
j_{cr}^{1}(t_{out})$. However, it maybe not so if the decoherence starts
earlier, for example, at the time instant $t_{dec}$, $t_{dec}<t_{out},$ and $%
t_{out}-t_{dec}$ is macroscopic; see discussion in the end of Section \ref%
{S2}. In this case, the quantity $\langle j^{1}(t_{out})\rangle$ decreases
significantly in low-dimensional systems, and increases in high-dimensional
systems.

We can compare contributions from the vacuum instability with contributions
from the vacuum polarization. Of course, all contributions due to pair
creation in expressions (\ref{emt-ren}) are exponentially small for the weak
electric field, $m^{2}/\left\vert qE\right\vert \gg 1$, so that the vacuum
polarization terms are principal. We are interested in the strong-field
limit, $m^{2}/\left\vert qE\right\vert \ll 1$. In such a limit, we obtain
from (\ref{p-emt}) that
\begin{align}
& \langle T_{00}\left( t\right) \rangle ^{p}=\langle T_{11}\left( t\right)
\rangle ^{p}\sim \left\vert qE\right\vert ^{d/2}\left\vert qE\right\vert
\left( t-t_{in}\right) ^{2},  \notag \\
& \langle T_{ii}(t)\rangle ^{p}\sim \left\vert qE\right\vert ^{d/2}\ln \left[
\sqrt{\left\vert qE\right\vert }\left( t-t_{in}\right) \right] ,\ \
i=2,3,\dots ,D,  \label{emt-cr}
\end{align}%
where the large dimensionless parameter $\sqrt{\left\vert qE\right\vert }%
\left( t-t_{in}\right) $ satisfies the stabilization condition (\ref%
{time-condition}). Comparing the evaluation of the EMT components from (\ref%
{emt-cr}) and (\ref{emt-lead}), we see immediately that, when $d$ is not a
multiple of four, the energy density of vacuum polarization, $\re \, \langle
T_{00}(t)\rangle _{ren}^{c}$, is negligible compared to the energy density
due to pair creation, $\langle T_{00}\left( t\right) \rangle ^{p}$, due to
inequality (\ref{time-condition}).

If $d=4$, the ratio $|\langle T_{00}(t)\rangle ^{p}\diagup \re \, \langle
T_{00}(t)\rangle ^{c}|$ in a massive theory with on-shell renormalization, $%
\mu =m$, is of the order $\left\vert qE\right\vert \left( t-t_{in}\right)
^{2}/\ln \left( \left\vert qE\right\vert /m^{2}\right)$. In order to
estimate the allowed values of the logarithm in the latter equation, we have
to have a physical model that describes the origin of the external classical
quasiconstant electric field. In problems of high-energy physics it is
usually assumed that just from the beginning there exists an uniform
classical electric field having a given energy density. The system of
fermions interacting with this field is closed, that is, the total energy of
the system is conserved. Under such an assumption, we take into account that
quantum electrodynamics in $d=4$ dimensions with the strong $T$-constant
external electric field can be considered as a consistent model only if the
backreaction due to pair creation is relatively small with respect to the
background, which implies the following restriction from above:%
\begin{equation}
\left\vert qE\right\vert ( t-t_{in})^{2} \ll \frac{\pi ^{2}}{2\alpha }\,,
\label{cons}
\end{equation}%
where $\alpha $ is the fine structure constant, see \cite{GG08-b,GG08-a}. We
consider macroscopic time intervals, such that $m^{2}\left( t-t_{in}\right)
^{2}\gg 1$, then it follows from (\ref{cons}) that $\left\vert qE\right\vert
/m^{2}\ll \pi ^{2}/\left( 2\alpha \right) $. Whence we obtain $\ln \left(
\left\vert qE\right\vert /m^{2}\right) \ll 6.5$. On the other hand,
according to condition (\ref{cons}), the maximum value allowed for $%
\left\vert qE\right\vert \left( t-t_{in}\right) ^{2}$ is two orders of
magnitude greater than the restriction obtained for the logarithm. Thus, we
see that in this case the quantity $\re \, \langle T_{00}(t)\rangle
_{ren}^{c}$ is negligible in comparison with $\langle T_{00}\left( t\right)
\rangle ^{p}$. This result can be generalized to $d=4n$ dimensions. Note
that the ratio of the transverse components of the pressure $|\langle
T_{ii}(t)\rangle ^{p}/\re\, \langle T_{ii}(t)\rangle ^{c}|$ is of the order $%
\ln \left[ \sqrt{\left\vert qE\right\vert }\left( t-t_{in}\right) \right] $
if $d\neq 4n$ and $\ln \left[ \sqrt{\left\vert qE\right\vert }\left(
t-t_{in}\right) \right] /\ln \left( \left\vert qE\right\vert /m^{2}\right) $
if $d=4n$. One can see, for example, in $d=4$ dimensions, that these
logarithms cannot be considered, in general, as really big quantities due to
restriction from above (\ref{cons}) and this evaluation can be generalized
to arbitrary dimensions. Then, in general, none of these transverse
components can be neglected. The evaluations may be different when another
model for the external classical quasiconstant field is considered, for
example, when there is an external source that supports a given external
field strength.

\section{Mean current and EMT in graphene\label{S4}}

\subsection{One-loop results in a given external field\label{SS4.1}}

It is known that at certain conditions electronic excitations in graphene
monolayer (just graphene, in what follows) behave as relativistic Dirac
massless fermions in $2+1$ dimensions. The so-called Dirac model for
electronic excitations in graphene was developed first by Semenoff in \cite%
{semenoff}, exploring results obtained decades earlier in the study of the
conductivity of graphite \cite{wallace}, see details in recent reviews \cite%
{castroneto,dassarma}. It was found that, at zero temperature and chemical
potential (i.e., at the so-called charge neutrality point), low-energy
electronic excitations in graphene are described in a tight-binding
approximation by the Dirac equation for massless particle in $2+1$
dimensions, with the Fermi velocity $v_{F}\simeq 10^{6}$ m/s playing the
role of the speed of light in relativistic particle dynamics. In this
section, we are going to explore such a correspondence, and consider
applications of the results obtained in the study of the quantized Dirac
field in an external background presented in the previous sections to some
problems of graphene physics that can be studied within the Dirac model. In
fact, we are going to study the electronic transport in graphene at low
carrier density and low temperatures when quantum interference effects are
important. Accordingly, we shall restrict from now on to the massless case
in $2+1$ dimensions.

First some comments about the Dirac model of graphene. There are actually
two species of fermions in this model, corresponding to excitations about
the two distinct Dirac points in the Brillouin zone of graphene, i.e., each
species belongs to a distinct valley. The algebra of $\gamma $-matrices has
two inequivalent representations in $(2+1)$-dimensions, as described in %
\eqref{eq:gamma-representations}, and a distinct (pseudo spin)
representation is associated with each Dirac point. There is no parity
anomaly in the Dirac model, in particular, the sum of current densities $%
\langle j^{\mu }(t)\rangle ^{c}$ (vacuum current contributions to the
probability amplitudes for processes with photons) for the two fermion
species, given by Eq.~(\ref{emt2a}), is zero. Note that the mean value of
transverse vacuum polarization current, given by Eq.~(\ref{emt2d}), is equal
to zero for each massless fermion species. For all other integral
quantities, since intervalley scattering can be neglected, the presence of
two valleys is taken into account simply by multiplying by the degeneracy
factor $2$.

Furthermore, there also is a spin degeneracy factor. The derivation of the
Dirac model starts from a nonrelativistic Schr\"{o}dinger equation for the
conduction electrons in graphene, leading to another doubling of fields due
to the (real) spin of the electron. As a result, there are four species of
fermions in the Dirac model  corresponding to graphene. The mean values that
we have obtained in the previous section hold true for each of the Dirac
fields independently. In order to find the corresponding mean values in
graphene, one should first add the contributions from the two valleys, as
discussed in the previous paragraph, and after that multiply by the spin
degeneracy factor two.

We consider an infinite flat graphene sample on which a uniform electric
field is applied, directed along the plane of the sample. We assume that the
applied field is the $T$-constant electric field studied in the previous
sections: the field is suddenly switched on at some time $t_{in}$, acting
then for a time-interval $T$, during which electron-hole pairs are created.
We consider the case of zero temperature and chemical potential, so that the
Dirac model can be used, and an initial state with neither electrons nor
holes.

Under these circumstances, the Eqs.~(\ref{n-cr}), (\ref{emt2c}), and (\ref%
{emt-ren}), multiplied by a degeneracy factor of four, describe,
respectively: $n_{g}^{cr}$, the total number density of electron-hole pairs
created by the electric field; $\langle j^{1}(t)\rangle _{g}$, the mean
longitudinal current density; and $\langle T_{\mu \mu }(t)\rangle _{g}$, the
mean EMT in the graphene. For the sake of comparison with known experimental
results, we are going to use SI units and restore the Planck constant $\hbar$
in this part of the work. We get the following results:
\begin{align}
& n_{g}^{cr}=r_{g}^{cr}T \, , \quad r_{g}^{cr}=\pi^{-2}\left(v_{F}\hbar
^{3}\right)^{-1/2} \left\vert eE\right\vert ^{3/2} \, ;  \label{g1} \\
& \langle j^{1}(t)\rangle _{g}=\mathrm{sgn}(E) A\Delta t \, , \quad
A=2ev_{F}r_{g}^{cr} \, ;  \label{g2} \\
& \langle T_{00}(t) \rangle _{g}=\langle T_{11}(t) \rangle_g = e \left\vert
E\right\vert v_{F}r_{g}^{cr}(\Delta t)^{2},  \notag \\
& \langle T_{22}\left( t\right) \rangle _{g}=\langle T_{22}(t) \rangle
_{g}^{p}+\langle T_{22}(t) \rangle _{g}^{c} \, ,  \notag \\
& \langle T_{22}(t) \rangle _{g}^{c}=\hbar \, r_{g}^{cr}\zeta _{R}(3/2) /2
\, ,  \notag \\
& \langle T_{22}(t) \rangle _{g}^{p}=\hbar \, r_{g}^{cr}\pi^{-1} \ln \left(
\sqrt{e\left\vert E\right\vert v_{F}/\hbar }\Delta t\right) \, ,  \label{g3}
\end{align}%
where $\Delta t=t-t_{in}$. These results hold true for all $t$ that satisfy
the stabilization condition (\ref{time-condition}), which has now the form:
\begin{equation}
\sqrt{e\left\vert E\right\vert v_F/\hbar} \Delta t \gg 1 \, .  \label{g4}
\end{equation}%
There appears a time scale specific to graphene,
\begin{equation}
\Delta t_{st}=\left( e\left\vert E\right\vert v_{F}/\hbar \right)^{-1/2} \, ,
\label{g4b}
\end{equation}
which plays the role of the stabilization time in the case under
consideration. The vacuum polarization contribution to the mean value $%
\langle T_{00}(t) \rangle _{g}=\langle T_{11}(t) \rangle _{g}$ in (\ref{g3})
is small due to Eq.~(\ref{g4}) and is neglected, that is, $\langle T_{00}(t)
\rangle _{g}=\langle T_{00}(t) \rangle _{g}^{p}$ and $\langle T_{11}(t)
\rangle _{g}=\langle T_{11}(t) \rangle _{g}^{p}$.

At $t\approx t_{out}$, we have $\Delta t\approx T=t_{out}$ $-t_{in}$ and
relations (\ref{mean-cr2}) show that the mean values $\langle
j^{1}(t)\rangle _{g}$ and $\langle T_{\mu \mu }\left( t\right) \rangle
_{g}^{p}$ hardly differ from the current density $j_{cr}^{1}(t_{out}$ $)$
and the quantity $T_{\mu \mu }^{cr}(t_{out})$ caused by created particles.
In the general case, the quantity $r_{g}^{cr}$ is the number density of
pairs of positive and negative charged states excited due to the constant
electric field per unit of time. Only at $t\approx t_{out}$, it can be
treated as the production rate of electron-hole pairs. In the presence of a
mass gap $\Delta \varepsilon =mv_{F}^{2}$, the rate $r_{g}^{cr}$ is
attenuated by a factor of $\exp \left[ -\pi (\Delta \varepsilon
)^{2}/e\left\vert E\right\vert v_{F}\hbar \right] $ according to Eq.~(\ref%
{n-cr}). In this case, the stabilization condition has general form (\ref%
{time-condition}) and the strong field condition reads $(\Delta \varepsilon
)^{2}/e\left\vert E\right\vert v_{F}\hbar \ll 1$.

There is a huge amount of papers on the conductivity in graphene, for the
most part on optical conductivity and on the minimal dc conductivity, see,
for example, a recent review of electronic transport in graphene  \cite%
{dassarma} and\emph{\ }an analysis of the situation with the minimal dc
conductivity in \cite{lewkowicz-10a,vildanov}. It is shown in \cite%
{lewkowicz-10a,lewkowicz-10b} that the time scale $\Delta t_{st}$ appears
for the tight-binding model as the time scale when the perturbation theory
with respect to electric field breaks down ($\Delta t_{st}\gg t_{\gamma }$,
where the microscopic time scale is $t_{\gamma }=\hbar /\gamma \simeq 0.24%
\mathrm{fs}$, with $\gamma =2.7$ eV being the hopping energy), and the dc
response changes from the linear in $E$ time-independent\emph{\ } regime to
a non-linear in $E$ and time-dependent regime. Thus, it was established that
the minimal dc conductivity occurs for ballistic flight times $\Delta t$
smaller than $\Delta t_{st}$. Our expression (\ref{g2}) is obtained for the
large time interval $\Delta t$ satisfying the condition ~(\ref{g4}). {Now
let us compare the results we have obtained with} the known results for
sufficiently large duration of electric field in the form of expression vs
expression.

The formula for the current density of created particles $j_{cr}^{1}(t_{out})
$ that follows from expression (\ref{g2}) agrees with the result obtained
from the WKB approach, see Eq. (8) in \cite{allor}, Eq\emph{s}. (20) and
(25) in \cite{dora}, and Eq.(A3) in \cite{vandecasteele} (the numerical
analysis of \cite{vandecasteele} performed in the frame of a non-equilibrium
Green function (NEGF) approach, which is referred to as a non-perturbative
quantum mechanical approach, is consistent with the semiclassical result).
In condensed-matter physics such a method is known as the Landau-Zener
approach (Note that the WKB approach is valid when the differential mean
number $\aleph _{\mathbf{p}}$, given by general expression (\ref%
{eq:np-formula}), is small. However, the WKB approximation for $\aleph _{%
\mathbf{p}}$ coincides with the exact expression (\ref{eq:Np-asymptotic}) in
the limit $T\rightarrow \infty $). The time-dependence of mean current
density $\langle j^{1}(t)\rangle _{g}$ given by Eq.~(\ref{g2}) is consistent
with the numerical solution of the first-quantized tight-binding model
equations obtained for the ballistic case in the time interval $\Delta
t_{st}<\Delta t<\Delta t_{B}$, see Eq.~(82) in Ref. \cite{lewkowicz-10b};
there a factor $3^{3/4}2^{-7/2}$ is replacing our $2/\pi ^{2}$, but these
factors are both equal to $0.20$, numerically. The $\Delta t_{B}=2\pi \hbar
(e\left\vert E\right\vert a)^{-1}$ is the Bloch time, required for the
constant electric field to shift the kinetic momentum across the Brillouin
zone ($a\approx 0.142\mathrm{\ nm}$ is the carbon-carbon distance) and, of
course, this value falls beyond to what may be covered by the continuous
Dirac model.

The expression for $\langle j^{1}(t)\rangle _{g}$ given by (\ref{g2}) is a
key formula in the study of the conductivity in the graphene at low carrier
density beyond the linear response in dc. It describes the mean electric
current of coherent carriers produced by the applied electric field. An
exotic feature of the electronic transport at the charge neutrality point,
as described by the Dirac model, is that one begins without any charges to
be accelerated at all: there are no electrons or holes initially. The
induced current can be considered, if one likes semiclassical style
comments, as a consequence of two mechanisms: charged pairs of
ultrarelativistic (with the constant velocity $v_{F}$) coherent electrons
and holes are first created with a strong suppression of large transverse
momenta $\left( \left\vert p_{2}\right\vert >\sqrt{e\left\vert E\right\vert
\hbar /v_{F}}\right) $ and after that their longitudinal kinetic momenta are
coherently increased by the electric field. The combined effect of both
processes that, in fact, cannot be localized and separated in the framework
of QED, is described by Eqs.~(\ref{g2}) and (\ref{g3}). In contrast to this,
at high carrier density one has to consider the increment of the
longitudinal kinetic momenta of an initially given number of incoherent
carriers by the electric field. In the latter case the transport theory
allows the semiclassical description of carriers.

It follows from our results that the mean current in the graphene is
parallel to the applied field, proportional to $\left\vert E\right\vert
^{3/2}$, and grows linearly with time. The fact that the current grows
indefinitely as $T\rightarrow \infty$ is a consequence of the absence of
scattering and a backreaction mechanisms in the model of unlimited size
under consideration: only effects caused by the applied external field have
been taken into account. In the experimental situation described in Ref.
\cite{vandecasteele}, a constant voltage between two electrodes connected to
the graphene was applied, and current-voltage characteristics ($I-V$) are
measured within $\sim 1$ s, which is a very large time-scale comparing with
the ballistic flight time $T_{bal}$ (the time which the electron spends to
cross the material of a finite length $L_{x}$), $T_{bal}=L_x/v_F$. To match
our results with these conditions, our time $T$ should be replaced by some
typical time-scale that we call the effective time duration $T_{eff}$. Some
kind of dissipation process may truncate the pair creation at $T_{dis}$, in
which case $T_{eff}=T_{dis}$. The standard candidates are collisions with
impurities, phonons, ripplons, and the electron-electron interactions. A
description of how the pair creation is counterbalanced by dissipative
processes so that a stable current settles down is still an open question.
In the absence of the dissipation, the transport is ballistic; in this case,
considering a strip with lateral infinite width and a finite length $L_x$,
we assume the ballistic flight time $T_{bal}$ to be the effective time
duration, $T_{eff}=T_{bal}$. It is experimentally shown, see \cite%
{dassarma,Bol08} and references therein, that the low temperature ($\Theta
\sim 5\mathrm{K}$) and gate voltage ($\left\vert V_{g}\right\vert <5$ V)
transport in current-annealed, suspended devices with a mobility $\sim
170000 \text{ cm}^2 \text{/Vs}$ is close to the ballistic limit over micron
dimensions. In a realistic sample, placed on a substrate, the effective time
duration $T_{eff}$ can be many times smaller than $T_{bal}$, because of
charged impurities or structural disorder of the substrate. However, such an
effective time $T_{eff}$ remains macroscopically large, so that Eq.~(\ref{g4}%
) still holds. One can expect that in realistic high-quality (high mobility)
samples the time $T_{eff}$ is comparable with the time $T_{bal}$. {In any
case, }for a finite flake length, the potential difference $V=EL_{x}$ is
finite, and one can consider the $I-V$ of graphene devices. Then Eq.~(\ref%
{g2}) describes a regime where the current behaves as $j\sim V^{3/2}$.

This fractional power dependence of the current-voltage characteristic in
graphene {is at present called} nonlinear (or superlinear) transport in
graphene ($I-V$ of the form $j\sim V^{3/2})$. In the Ref.~ \cite%
{allor,dora,lewkowicz-10b}, this behavior was related to a possible pair
creation by a constant electric field and experiments aiming at the
observation of the effect were proposed. An experimental observation was
recently reported \cite{vandecasteele}. It is agreed that such a superlinear
transport is a distinctive feature characterizing the regime dominated by
pair creation.

It should be noted that our description of the quantum transport in graphene
in the framework of strong-field QED is not restricted by a semiclassical
approximation of carriers and it does not use any statistical assumptions
inherent in the Boltzmann transport theory. The estimations that follow show
that this is important for the study of the conductivity close to the Dirac
point. A typical density $n_{g}^{cr}$ of carriers created in the ballistic
case, given by Eq.~(\ref{g1}), e.g. for typical $V\sim 1$ V and $L_x \sim 1$
$\mathrm{\mu m}$, is of the order $n_{g}^{cr} \sim 6 \times 10^{11}$ $%
\mathrm{cm}^{-2}$. Thus, in the general case, the density $n_{g}^{cr}$ is of
the same order as the impurity density $n_{imp}$ ($10^{10}$-$10^{12}\mathrm{%
cm}^{-2}$ \cite{dassarma}). It should be stressed that this estimate is made
for the ballistic case when $T_{eff}=T_{bal}$. If $T_{eff}\ll T_{bal}$, then
the density $n_{g}^{cr}$ is less than $n_{imp}$. Moreover, the electric
field creates the carriers in pure states with distribution (\ref%
{eq:Np-asymptotic}) that differs significantly from the equilibrium
distribution. The carriers created at relatively small times $T_{eff}\sim
\Delta t_{st}$ are wave packets with partial plane waves, which have wide
range of wave lengths from $\infty $ to approximately $0.2$ $\mathrm{\mu m}$%
. These lengths correspond to maximum kinetic momenta of the order $\sqrt{%
e\left\vert E\right\vert \hbar /v_{F}}$, see (\ref{eq:Np-asymptotic}). As $%
T_{eff}$ increases the range of the longitudinal kinetic momenta of created
particle grows to its maximum $e\left\vert E\right\vert T_{eff}\gg \sqrt{%
e\left\vert E\right\vert \hbar /v_{F}}$; at the same time the range of their
transverse kinetic momenta remains unchanged. Thus, in the ballistic case,
the lower bound of the range of the longitudinal wavelengths decreases
achieving its minimum of the order $4$ $\mathrm{nm}$. The estimated minimum
wavelength of the order $0.2$ $\mathrm{\mu m}$ is not much less than sample
sizes that were used in experiments, and in some cases they are of the same
order. Thus, we see that the Boltzmann theory, based on WKB approximation,
is not well adapted to describe evolution of massless carries created from
the vacuum. This is a nonstandard situation for usual transport problems in
condensed matter physics. Our approach gives a non-perturbative description
of the system evolution when quantum interference effects are important. We
see that the QED derivation of the superlinear $I-V$ presented above gives
additional arguments in favor of the interpretation that such a behavior is
due to the pair creation. At the same time, in order to continue the study
of the conductivity of graphene in the framework of the QED methods, one has
to analyze in detail physical conditions in the graphene under which all the
machinery, and especially Eqs.~(\ref{g2}) and (\ref{g3}), are valid.

\subsection{Mean electromagnetic field\label{SS4.2}}

The mean time-dependent current $\langle j^{1}(t)\rangle _g$ is the source
in the Maxwell equations for a mean electromagnetic field $( \mathbf{\bar{E},%
\bar{B}})$, where $\mathbf{\bar{E}=E}+\mathbf{E}_{rad}$ and $\mathbf{\bar{B}%
=B}_{rad}$ are electric and magnetic components, respectively. Here, the
initial external constant uniform electric field $\mathbf{E}$ satisfies the
homogeneous Maxwell equations, and the fields $\mathbf{E}_{rad}$ and $%
\mathbf{B}_{rad}$ are due to the current $\langle j^{1}(t)\rangle _{g}$,
representing the backreaction of created pairs to the external field. The
charged fermions in the graphene should feel the total field $\left( \mathbf{%
\bar{E},\bar{B}}\right)$; in particular, the pair creation is induced by the
mean electric field $\mathbf{\bar{E}}$. Since the current $\langle
j^1(t)\rangle _{g}$ increases linearly as $\Delta t$ increases, the
backreaction to the external field should become relevant at some time
instant. The situation looks similar to that studied in \cite{GG08-b,GG08-a}
for QED in $3+1$ dimensions, although there are considerable peculiarities
in the case under consideration. Dealing with graphene devices, it is
natural to assume that the constant strength $E$ on the graphene plane is
due to the applied fixed voltage $V$ and therefore is not changed when the
created charges flow into the reservoirs which are located outside the
graphene and have sufficiently large capacitances. It is assumed that an
external current flows to the electrodes to maintain the fixed voltage,
i.e., we are dealing with an open system of fermions interacting with
classical electromagnetic field. Then Maxwell equations can be solved
through infinite space and we may assume that the strength $E$ is fixed by
boundary condition at the infinity. The electromagnetic field is not
confined to the graphene surface, $z=0$, but rather propagates in the
ambient $3+1$ dimensional space-time, where $z$ is the coordinate of axis
normal to the graphene plane. To take this into account, we have to present
the current term in the Maxwell equations as the current $\mathbf{J=}( J,0,0)
$ restricted to a plane immersed in the $3$-dimensional space. We are
interested in the time intervals $\Delta t$ that satisfy the inequality $%
0\leq \Delta t\leq T$. Then, taking into account that the current $\langle
j^1(t)\rangle_g$ appears for $\Delta t>0$, we can write $J=\langle
j^1(t)\rangle_g \, \delta (z) \theta (\Delta t)$ for $\Delta t \leq T$. The
nontrivial Maxwell equations to be solved are:
\begin{equation}
\boldsymbol{\nabla} \times \mathbf{\bar{B}} = \mu_0 \mathbf{J} + \frac{1} {%
c^{2}} \frac{\partial \mathbf{\bar{E}}}{\partial t} \, , \quad \boldsymbol{%
\nabla} \times \mathbf{\bar{E}} = - \frac{\partial \mathbf{\bar{B}}}{%
\partial t} \, ,  \label{g5}
\end{equation}
where $\mu _{0}$ is the magnetic permeability and $c$ is the speed of light
in the vacuum.

The only nonzero components of the irradiated field are the $x$-component of
the electric field, $E_{x}^{rad}$, and the $y$-component of the magnetic
field, $B_y^{rad}$. Solving the Maxwell equations under initial condition $%
E_{x}^{rad}=0$ and $B_{y}^{rad}=0$ at $\Delta t=0$, one finds that for $%
0\leq \Delta t\leq T$, the electromagnetic field produced by the current
confined to the graphene sheet is:
\begin{align}
& E_x^{rad}=-\frac{\mu_0}{2}\mathrm{sgn} (E) A ( c\Delta t-\left\vert
z\right\vert) \theta (c\Delta t-|z|) \, ,  \notag \\
& B_y^{rad}=\mathrm{sgn} (z) E_x^{rad}/c \, .  \label{g6}
\end{align}

The direction of the induced magnetic field changes sign across the graphene
sheet, and $\mathrm{sgn}\left( z\right) =\pm 1$ corresponds to upper and
lower regions. The electric field $E_{x}^{rad}$ is opposite to the applied
field $E$ and is continuous in $z$-direction, in particular, at $z=0$. A
real graphene flake is a film of very large dimension in the $x-y$ plane and
a finite monolayer atomic thickness of approximately  $0.1\mathrm{nm}$ to $%
0.2\mathrm{nm}$ in the $z$-direction \cite{dassarma}. For realistic densities of carriers
the potential field in $z$-direction looks like a deep potential well, which
effectively forbids the motion of carries in this direction.\emph{\ }Then it
is natural to assume that the induced electromagnetic field inside the
graphene is the mean between limiting values of the field from upper and
lower regions as $\left\vert z\right\vert \rightarrow 0$. Thus, the
intensity of the induced electric field on the graphene plane reads:
\begin{equation}
E_{x}^{rad}=-\mathrm{sgn}\left( E\right) \frac{4\alpha }{\pi }\left\vert
E\right\vert ^{3/2}\sqrt{v_{F}e/\hbar }\,\Delta t\,,  \label{g7}
\end{equation}%
where $\alpha =\mu _{0}ce^{2}/2h$ is the fine structure constant. As to the
induced magnetic field, we believe that it is zero inside the graphene.

The QED with an external constant electric field is a consistent model as
long as the field produced by the induced current is negligible compared to
the applied field, $\left\vert E_{x}^{rad}\right\vert \ll \left\vert
E\right\vert $. {This gives the consistency restriction:}
\begin{equation}
\sqrt{v_{F}e\left\vert E\right\vert /\hbar }\,\Delta t\ll \frac{\pi }{%
4\alpha }\,.  \label{g8}
\end{equation}%
In this case, the external electric field can be considered as a good
approximation of the effective mean field. We call the typical time scale
related to Eq.~(\ref{g8}), $\Delta t_{br}=\Delta t_{st}\pi /4\alpha $, the
time of backreaction. On the other hand, the dimensionless parameter in the
l.h.s. of Eq.~(\ref{g8}) satisfies the stabilization condition given by Eq.~(%
\ref{g4}). Thus, there is a {window} in the parameter range where the model
is consistent, $t_{\gamma }\ll \Delta t_{st}\ll \Delta t_{br}$. Moreover,
this restriction corresponds to a specific regime which might be relevant to
some known experiments with graphene, as we shall see below.

In the description of the carrier creation inside a graphene flake of a
finite length $L_{x}$, we use a simple picture of the pair creation due to
the $T$-constant field. We assume that the maximal duration of the electric
field in our model is the effective time duration $T_{eff}$, {and that, in
the case of ballistic transport, $T_{eff}=T_{bal}$}. In typical experiments,
$L_{x}\sim 1\mathrm{\mu m}$, and then $T_{bal}\sim 10^{-12}\mathrm{s}$.
Taking $\Delta t=T_{bal}$ in Eqs.~(\ref{g4}) and (\ref{g8}), we obtain the
following restrictions on the electric field:
\begin{equation}
7\times 10^{2}\mathrm{V/m}\ll \left\vert E\right\vert \ll 8\times 10^{6}%
\mathrm{V/m}\,.  \label{g9}
\end{equation}%
Since the voltage is $V=EL_{x}$, one finds the inequalities
\begin{equation}
7\times 10^{-4}\,\mathrm{V}\ll V\ll 8\,\mathrm{V}\,.
\label{eq:voltage-bounds}
\end{equation}%
These voltages are in the range typically used in experiments with graphene.

In the experiment described in \cite{vandecasteele}, for instance, the $I-V$
curve was studied in a range from decivolts to a few volts, in samples with
lengths $L_{x}$ varying from $0.9$ to $5.9$ $\mathrm{\mu m}$ and widths from
$70$ to $1500$ $\mathrm{nm}$. A power law $I\sim V^{\delta }$ was used to
fit the $I-V$ curves near the Dirac point, and it was found that $1\leq
\delta \leq 3/2$. It was shown that the $I-V^{\prime }$s in the graphene
devices become superlinear in the presence of disorder (in low mobility
samples) while in high-quality (high mobility) samples, the superlinearity
is masked by some other effect. The exponent $\delta $ looks like a
monotonically decreasing function of mobility for different devices and
superlinearity vanishes for devices with high mobility, $\delta \rightarrow 1
$. These results were interpreted as an interplay between pair creation and
scattering by charged impurities and optical phonons. According to this
interpretation, the superlinearity of the $I-V$ in low mobility samples is
in a qualitative agreement with expression (\ref{g2}), while it is
compensated by the contribution of the intraband current (the current of
carriers that were present before the electric field was switched on, due to
an imperfect experimental realization of the Dirac point conditions), which
tends to saturate in high mobility samples due to interaction with optical
phonons. The absence of a compensation of the superlinearity in low mobility
samples is explained by the fact that the {presence of a large number of
charged impurities prevents the growth of the intraband current, making the
interaction with optical phonons irrelevant. Except for this qualitative
analysis, however, there is no theoretical description of these
observations; the numerical analysis (in the NEGF approach) of \cite%
{vandecasteele}, in particular, does not include the optical phonons which
are central to their interpretation of the effect, and it is argued that
other effects should also be relevant for the analysis, as self-heating of
the graphene sample, for instance \cite{vandecasteele}. }

In fact, according to our analysis, backreaction should not be neglected. We
have shown that the consistency of the Dirac model with a given external
field is {ensured} under the conditions (\ref{g4}) and (\ref{g8}) for $%
\Delta t$ and $E$, which are thus extremely important for obtaining adequate
physical results in course of the calculations. Analyzing the experimental
settings one can be sure that the stabilization condition \eqref{g4} is
satisfied for all known measurements, which means that the lower bound in %
\eqref{eq:voltage-bounds} is respected. However, the condition (\ref{g8})
could be violated because the voltage varies from $1$ to $5\,\mathrm{V}$,
and sample length varies from $0.9$ to $5.9$ $\mathrm{\mu m}$, as reported
in \cite{vandecasteele}. Therefore, backreaction (appearance of the induced
electric field $E_{x}^{rad}$) cannot be ignored in calculating the mean
current in high mobility samples. This means that, in the ballistic case,
Eq.~(\ref{g2}) can give overestimated values for the current density of
created particles for a given voltage. In the case of low mobility samples,
the effective duration time $T_{eff}$ is due to the scale of dissipation
processes $T_{dis}$, which can be many times less than the time $T_{bal}$,
so that condition (\ref{g8}) holds true at $\Delta t=T_{dis}$, and the
regime of backreaction is not reached. As mobility is increased,
backreaction becomes more pronounced, and it might compensate the
superlinearity of the pair-creation process. In order to investigate whether
this effect can lead to a transition from a superlinear to a linear $I-V$
curve, we investigate in the next section how backreaction affects
pair-production in graphene.

\subsection{Effective mean field and mean current\label{SS4.3}}

In this section, we study a possible generalization of the above considered
model with the $T-$constant external field. In particular, in the framework
of this generalization we are going to take into account the backreaction of
the mean current to the applied electric field.

Let us consider time intervals when the external electric field is switched
on, $0\leq \Delta t\leq T$. In realistic cases, after switching on, the
external field $E$ remains constant on the graphene plane due to the applied
voltage $V$, which is supported by external sources to remain fixed.
Accordingly, we assume that on the graphene plane $E$ cannot be changed due
to pair creation inside of the sample. As to the electric field $\mathbf{%
\bar{E}}$ inside the graphene, we suppose that it can vary with time and it
is directed along the axis $x$, so that $\mathbf{\bar{E}}(t) =(\bar{E}%
(t),0,0)$. The $x$-component $\bar{E}\left( t\right) $ is a superposition of
the applied external field $E$ and a time-dependent electric field $%
E_{x}^{rad}(t)$ irradiated by the current induced in the sample, $\bar{E}%
(t)=E+E_{x}^{rad}(t)$. In the initial time instant $t_1=t_{in}$, we have $%
\bar{E}(t_1)=E$. Let $\bar{E}(t)$ be a slowly varying field within a time
interval $\Delta t_i=t_{i+1}-t_{i}>0$ such that the following condition
holds:
\begin{equation}
\left\vert \frac{1}{\bar{E}(t)} \frac{\partial \bar{E}(t)} {\partial t}%
\right\vert \Delta t_i \ll 1 \, , \quad t \in (t_i,t_{i+1}] \, ,  \label{g10}
\end{equation}
and let $\Delta t_i$ be large enough to obey the stabilization condition of
the type (\ref{g4}),
\begin{equation}
\sqrt{e\left\vert \bar{E}(t) \right\vert v_F/\hbar }\Delta t_i \gg 1 \, ,
\quad t\in (t_i,t_{i+1}] \, .  \label{g11}
\end{equation}
We consider now the time evolution of the mean field and the vacuum mean
current between the initial time $t_{in}$ when the external field is
switched on and a time instant $t_{fin}$, which coincides with $t_{out}$, or
precedes it, $t_{fin}\leq t_{out}$. We first divide the interval $%
(t_{in},t_{fin}]$ into $N$ equal intervals $\Delta t_{i}$, such that $\Delta
t_1=\Delta t_2=\cdots =\Delta t_N$, $\sum_{i=1}^{N}\Delta t_i=t_{fin}-t_{in}$%
. We suppose that Eqs.~(\ref{g10}) and (\ref{g11}) hold true for all
intervals. That allows us to treat the electric field as approximately
constant within each interval, $\bar{E}(t)\approx \bar{E}(t_i)$, for $t\in
(t_{i},t_{i+1}]$.

Let us find the vacuum mean current density for the slowly varying electric
field $\bar{E}\left( t\right) $. We begin with the case $t\in (t_1,t_2]$.
Here, according to Eqs.~(\ref{Delta-eff}) and (\ref{j-GF}), the current
density $\langle j^{1}(t)\rangle _{g}$ is formed by the contributions from
the vacuum states excited by the field $E$, these contributions having
momenta that are restricted to the region (\ref{domain-def}). The phase
volume of this region depends both on the magnitude of the electric field $E$
and on the interval $t-$ $t_{1}$. It follows from the Eq.~(\ref{Delta-eff})
that the density $\langle j^{1}(t)\rangle _{g}$ grows as $t-t_1$ because the
field $E$ excites additional states with larger longitudinal momentum $p_1$.
In the end of this time interval, when $t=t_2$, the current density $\langle
j^{1}(t)\rangle_g$ takes the form:
\begin{equation}
\Delta j_1 = \mathrm{sgn}(E) D\left\vert \bar{E}(t_1)
\right\vert^{3/2}\Delta t_1 \, ,  \label{g12}
\end{equation}
with $D=2\pi^{-2} v_{F}^{1/2} \hbar^{-3/2}e^{5/2}$. According to Eq.~(\ref%
{g7}), the field $E_x^{rad}(t)$ has a direction opposite to $\bar{E}\left(%
\bar{t}_1\right)=E$, and from inequality (\ref{g10}) its magnitude is small
in comparison with it, $|E_{x}^{rad}(t)|\ll |\bar{E}(\bar{t}_{1})|$.
Therefore, Eq.~(\ref{g12}) determines the leading term in the mean value $%
\langle j^{1}(t)\rangle_g$ in the corresponding large $\tau$-limit and in
the mean field approximation, and any corrections to $\langle
j^{1}(t)\rangle_g$ due to derivative $\partial_t \bar{E}(t)$ are small. One
can see that $\left\vert \bar{E}(t_2) \right\vert < \left\vert \bar{E}(t_1)
\right\vert$. Now let us proceed to the case $t\in (t_{2},t_{3}]$. The
effective mean field $\bar{E}(t)$ is already different from $E$ and, as
before, the field $\bar{E}(t)$ is approximately constant, $\bar{E}(t)\approx
\bar{E}(t_2)$.

The constant field $\bar{E}(t_2)$ acting on the interval $(t_2,t_3]$ excites
new vacuum states which do not participate in the formation of the current
density $\langle j^{1}(t)\rangle _{g}$ on the previous step. These new
states have longitudinal momenta $p_1 \mathrm{sgn} (E)$ greater than states
excited from the vacuum during the time $t_2-t_1$ and transversal momenta $%
p_2$ limited by the field $\bar{E}(t_2)$. The region of these momenta is
defined by inequalities similar to those in (\ref{domain-def}). Such
inequalities in the mean field approximation have the form
\begin{align}
& |p_2|\leq \sqrt{e|\bar{E}(t_i)| \hbar /v_{F}}\left[ \sqrt{e |\bar{E}(t_i)|
v_F/\hbar} \, (t-t_i) -K\right]^{1/2} ,  \notag \\
&t_{i} + \frac{K}{\left\vert e\bar{E}(t_i) v_F/\hbar \right\vert^{1/2}} \,
\leq -\frac{p_1}{e\bar{E}(t_i)} \leq t-\frac{K}{\left\vert e\bar{E}(t_i)
v_F/\hbar \right\vert^{1/2}} \, .  \label{g13}
\end{align}
In the case $t\in (t_{2},t_{3}]$, one has to set $t_{i}=t_{2}$ in these
relations. As a result, the current density $\langle j^{1}(t)\rangle _{g}$
in the second interval takes the form
\begin{align}
& \langle j^{1}(t)\rangle_g=\Delta j_{1}+\Delta j(t-t_2) \, ,  \notag \\
& \Delta j(t-t_2) = \mathrm{sgn} (E) D\left\vert \bar{E}(t_2)
\right\vert^{3/2}(t-t_2) \, ,  \label{g14}
\end{align}
in the large $\tau $-limit. As in the case considered above, this current
irradiates the corresponding field $E_{x}^{rad}$, which is directed against
to the constant field $\bar{E}(t_2)$, but its magnitude is less than the
latter. Because $t-t_{2}\leq \Delta t_{2}$ and it is supposed that condition
(\ref{g10}) holds true, one can see that the contribution from the
derivative $\partial_{t}\bar{E}(t)$ to the current density $\langle
j^{1}(t)\rangle_{g}$ is much less than the leading term, given by (\ref{g14}%
). Due to the contribution $E_{x}^{rad}$, the mean field will decrease, $%
\left\vert \bar{E}(t) \right\vert < \left\vert \bar{E}(t_2) \right\vert$,
reaching its minimal value at $t=t_{3}$. The further evolution of the vacuum
mean current density and mean field has similar behavior for time $t$ from
any interval $\Delta t_{i}$. In each such intervals the region of the
momenta is given by Eq.~(\ref{g13}). In the general case (in the
corresponding large $\tau $-limit) when $t\in (t_{M},t_{M+1}]$, $2\leq M\leq
N$, we can represent the current density $\langle j^{1}(t)\rangle _{g}$ as
the following sum of partial contributions:
\begin{align}
&\langle j^{1}(t)\rangle _{g}=\sum_{i=1}^{M-1}\Delta j_{i}+\Delta j(t-t_M)
\, ,  \notag \\
&\Delta j_{i}=\mathrm{sgn}(E) D\left\vert \bar{E}(t_i) \right\vert^{3/2}
\Delta t_i \, ,  \notag \\
& \Delta j(t-t_M) =\mathrm{sgn}(E) D\left\vert \bar{E}(t_M) \right\vert
^{3/2}(t-t_M) \, .  \label{g15}
\end{align}
Condition (\ref{g10}) guarantees that corrections to the leading terms $%
\Delta j_{i}$ and $\Delta j(t-t_M)$, given by (\ref{g15}), due to the
derivatives $\partial_t \bar{E}(t)$ are small in each interval $%
(t_{i},t_{i+1}]$. In each time interval, the mean current irradiates the
corresponding field $E_{x}^{rad}$ directed against the external field $E$,
thus, the mean field $\bar{E}(t)$ is a monotonically decreasing function for
$t\in (t_{in},t_{fin}]$.

Let us use the approximations
\begin{align*}
& \left\vert \bar{E}(t_i) \right\vert ^{3/2}\Delta t_{i}\approx
\int_{t_{i}}^{t_{i+1}}\left\vert \bar{E}(t) \right\vert^{3/2}dt \, , \\
& \left\vert \bar{E}(t_M) \right\vert^{3/2}(t-t_M) \approx
\int_{t_{M}}^{t}\left\vert \bar{E}\left( \tilde{t}\right) \right\vert ^{3/2}d%
\tilde{t} \, .
\end{align*}
Then the mean value $\langle j^{1}(t)\rangle _{g}$ can be approximated by
the following integral form:
\begin{equation}
\langle j^{1}(t)\rangle_g = \mathrm{sgn}(E) D\int_{t_{in}}^{t}\left\vert
\bar{E}( \tilde{t}) \right\vert^{3/2} d\tilde{t} \, .  \label{g16}
\end{equation}
Note that Eq.~(\ref{g16}) holds true within large $\tau$-limit, in which we
neglect a contribution from the derivative $\partial_t\bar{E}(t)$, if the
condition (\ref{g10}) is valid for all the intervals $(t_{i},t_{i+1}]$.

In a similar manner, one can find a unified representation of the vacuum
mean current density $\langle j^{1}(t)\rangle _{g}$ and the diagonal
elements $\langle T_{\mu \mu }\left( t\right) \rangle _{g}^{p}$ for the
slowly varying electric field $\bar{E}\left( t\right)$ in the same
approximation:
\begin{widetext}
\begin{align}
& \langle j^{1}(t)\rangle _{g}=i8e\left. \bar{P}_{1}\bar{\Delta}%
^{p}(x,x^{\prime })\right\vert _{x=x^{\prime }} \, , \quad \langle T_{\mu \mu
}(t)\rangle _{g}^{p}=i8\left. \bar{P}_{\mu }^{2}\bar{\Delta}^{p}(x,x^{\prime
})\right\vert _{x=x^{\prime }} \, ,  \label{g26} \\
& \bar{\Delta}^{p}(x,x^{\prime })=-i\int_{t_{in}+K\Delta t_{st}}^{t-K%
\overline{\Delta t}_{t}}\bar{h}_{\perp }(\mathbf{x}_{\perp },\mathbf{x}%
_{\perp }^{\prime })\bar{h}_{\parallel }(x_{\parallel },x_{\parallel
}^{\prime })d\tilde{t}\,,  \notag \\
& \bar{h}_{\parallel }(x_{\parallel },x_{\parallel }^{\prime })=\frac{%
\left\vert \bar{E}\left( \tilde{t}\right) \right\vert }{\left\vert \bar{A}%
_{1}(t)\right\vert -\left\vert \bar{A}_{1}(\tilde{t})\right\vert }\,\exp %
\left[ -\frac{i}{\hbar }e\bar{A}_{1}(\tilde{t})\left( x_{1}-x_{1}^{\prime
}\right) \right] \cos \left\{ \frac{1}{2}\left[ \bar{\xi}(t^{\prime })^{2}-%
\bar{\xi}(t)^{2}\right] \right\} ,  \notag \\
& \bar{h}_{\perp }(x_{\perp },x_{\perp }^{\prime })=\left( \frac{v_{F}}{%
\hbar ^{3}}\right) ^{1/2}\frac{\left\vert e\bar{E}\left( \tilde{t}\right)
\right\vert ^{1/2}}{(2\pi )^{2}}\exp \left[ -\frac{\left(
x_{2}-x_{2}^{\prime }\right) ^{2}e\left\vert \bar{E}\left( \tilde{t}\right)
\right\vert }{4\pi \hbar v_{F}}\right] \,,  \label{g27}
\end{align}
\end{widetext}
where
\begin{align*}
&\bar{P}_{0}=i\frac{\hbar }{v_{F}}\frac{\partial }{\partial t} \, , \quad
\bar{P}_1=i\hbar \frac{\partial }{\partial x^{1}}+e\bar{A}_{1}(t) \, , \quad
\bar{P}_2=i\hbar \frac{\partial }{\partial x^{2}} \, , \\
&\frac{\hbar }{v_{F}}\bar{\xi}(t)\frac{d\bar{\xi}(t)}{dt}=e\left[ \left\vert
\bar{A}_{1}(t)\right\vert -\left\vert \bar{A}_{1}(\tilde{t})\right\vert %
\right] \, ,  \notag \\
& \bar{A}_{1}(t)=\int_{t_{in}}^{t}\bar{E}\left( \tilde{t}\right) d\tilde{t}%
+Et_{in} \, ,
\end{align*}
$\overline{\Delta t}_{t}=\left( e\left\vert \bar{E}(t) \right\vert
v_{F}/\hbar \right)^{-1/2}$, and $\Delta t_{st}$ is the characteristic
time-interval determined by Eq.~(\ref{g4b}). These equations are a
generalization of representation (\ref{j-GF}) for the case under
consideration, where $\bar{\Delta}^{p}(x,x^{\prime })$ is a generalization
of the function $\Delta^{p}(x,x^{\prime })$ in (\ref{Delta-eff}). One can
see that the expression $\langle j^{1}(t)\rangle_g$ given by (\ref{g26})
coincides with that given by Eq.~(\ref{g16}) in the large $\tau$-limit. In
this approximation the mean values $\langle T_{\mu \mu}(t) \rangle_g$ have
the form:
\begin{align}
& \langle T_{00}\left( t\right) \rangle _{g} = \langle T_{11}(t) \rangle
_{g}=\langle T_{11}(t) \rangle _{g}^{p} \, ,  \notag \\
& \langle T_{22}(t) \rangle_{g}=\langle T_{22}(t) \rangle _{g}^{p}+\langle
T_{22}(t) \rangle_g^c \, ,  \notag
\end{align}
with
\begin{align}
\langle T_{11}(t) \rangle _{g}^{p} = & 2ev_{F}\rho \int_{t_{in}+K\Delta
t_{st}}^{t-K\overline{\Delta t}_{t}}\left\vert \bar{E} \left( \tilde{t}%
\right) \right\vert ^{3/2}  \notag \\
& \quad \times \left[ \left\vert \bar{A}_{1}(t)\right\vert -\left\vert \bar{A%
}_{1}(\tilde{t})\right\vert \right] d\tilde{t} \, ,  \notag \\
\langle T_{22}(t) \rangle _{g}^{p} = & \pi ^{-1}\hbar \rho
\int_{t_{in}+K\Delta t_{st}}^{t-K\overline{\Delta t}_{t}}\frac{\left\vert
\bar{E}\left( \tilde{t}\right) \right\vert ^{5/2}}{\left\vert \bar{A}%
_{1}(t)\right\vert-\left\vert \bar{A}_{1}(\tilde{t})\right\vert }d\tilde{t}
\, ,  \notag \\
\langle T_{22}(t) \rangle _{g}^{c} = & \frac{1}{2}\zeta _{R}\left(
3/2\right) \hbar \rho \left\vert \bar{E}\left( t\right) \right\vert ^{3/2}
\, ,  \label{g28}
\end{align}
and $\rho =\pi^{-2} v_{F}^{-1/2} \hbar^{-3/2} e^{3/2}$.

By using Eq.~(\ref%
{g16}), we obtain that in the Maxwell equations for the mean electromagnetic
field the current term for $t\leq t_{fin}$ has the form $J=\langle
j^{1}(t)\rangle _{g}\delta(z) \theta (t-t_1)$. The only nonzero components
of the radiated field are the $x$-component $E_{x}^{rad}$ of the electric
field and the $y$-component $B_{y}^{rad}$ of the magnetic field. Solving the
Maxwell equations (\ref{g5}) under the initial conditions $E_{x}^{rad}=0$
and $B_{y}^{rad}=0$ at $t=t_{in} $, one finds that for $0\leq t-t_{in}\leq
t_{fin}-t_{in}$, and within a distance {$|z|$} smaller than $c\left(
t-t_{in}\right) $ to the plane, the electromagnetic field produced by the
current $\langle j^{1}(t)\rangle _{g}$ confined to the graphene sheet is:
\begin{align}
E_{x}^{rad}(t,z) = & -\mathrm{sgn}(E) \frac{\mu _{0}D}{2}\int \left\vert
\bar{E}(t-l/c) \right\vert ^{3/2}  \notag \\
& \quad \times \theta( t-t_{in}-l/c) \theta (l-|z|) dl \, ,  \notag \\
B_{y}^{rad}(t,z) = & \mathrm{sgn}(z) E_{x}^{rad}(t,z) /c \, .  \label{g17}
\end{align}
It can be seen that the radiated field is a function of only the light-cone
variable, $E_{x}^{rad}(t,z) = E_{x}^{rad}(t-\left\vert z\right\vert /c)$.

The electric field $E_{x}^{rad}(t,z)$ has the same limit value near the
plane from the upper and lower regions and we assume that $E_{x}^{rad}(t,z)$
is continuous at $z=0$. Thus, inside the graphene film the irradiated
electric field is $E_{x}^{rad}(t) =E_{x}^{rad}(t,0)$. The mean magnetic
field inside the graphene film is zero. Extracting the derivative $\partial
_{t}\bar{E}(t) =\partial _{t}E_{x}^{rad}(t)$ from (\ref{g17}), we get that
for $t\in (t_{in},t_{fin}]$ a self-consistency equation for the mean field
inside the graphene has the form:
\begin{equation}
\frac{d\left\vert \bar{E}(t) \right\vert }{dt}=-\frac{4\alpha }{\pi} \sqrt{%
\frac{ev_F}{\hbar} }\left\vert \bar{E}(t) \right\vert ^{3/2} \, , \quad t
\in (t_{in},t_{fin}] \, .  \label{g18}
\end{equation}
As was already established, the mean field field $\bar{E}(t)$ is collinear
to the external field $E$. Taking into account the initial condition, $\bar{E%
}(t_{in})=E$, one finds the solution of equation (\ref{g18}) as follows:
\begin{equation}
\bar{E}(t) = E \epsilon(t)^{-2} \, , \quad \epsilon(t) = \left[ 1+\frac{%
2\alpha}{\pi} (t-t_{in}) /\Delta t_{st} \right] \, .  \label{g19}
\end{equation}
Then the irradiated field is $E_{x}^{rad}(t)=\bar{E}(t)-E$.

As $t\rightarrow t_{fin}$ the magnitude of the mean field $\bar{E}(t)$
decreases. Therefore, condition (\ref{g11}) will be satisfied for all the
intervals $(t_{i},t_{i+1}]$ if it is satisfied for the last interval $%
(t_{N},t_{fin}]$. In our approximation, this condition defines the minimal
possible length of the intervals $\Delta t_{i}$,
\begin{equation}
\Delta t_{i}\gg \overline{\Delta t}=\Delta t_{st}\epsilon(t_{fin}) \, .
\label{g20}
\end{equation}
Condition (\ref{g10}) is satisfied for all the intervals $(t_{i},t_{i+1}]$,
if it is satisfied for the period $(t_{1},t_{2}]$. With $\Delta t_{1} \gg
\overline{\Delta t}$, the latter implies the following restriction for the
interval $t_{fin}-t_{in}$:
\begin{equation}
\frac{4\alpha }{\pi }\epsilon(t_{fin}) \ll 1 \Longrightarrow
(t_{fin}-t_{in}) /\Delta t_{st} \ll \frac{\pi ^{2}}{8\alpha ^{2}} \, .
\label{g22}
\end{equation}%
The mean field approximation is consistent if inequality (\ref{g22}) holds
true. This restriction is much weaker than the one given by Eq.~(\ref{g8}),
which ensures consistency of the external field approximation. Thus, in the
approach under consideration the {largest} time scale is $\Delta t_{fin}$,
which is related to time scales considered before as follows:
\begin{equation*}
\Delta t_{fin}=\frac{\pi }{2\alpha }\Delta t_{br}=\frac{\pi ^{2}}{8\alpha
^{2}}\Delta t_{st} \, .
\end{equation*}
When $t_{fin}-t_{in}$ approaches $\Delta t_{fin}$, the magnitude $\left\vert
\bar{E}(t_{fin}) \right\vert$ decreases and becomes of the order $\left\vert
E_{\min }\right\vert =\left\vert E\right\vert (4\alpha /\pi)^{2}$. One can
verify that in the case of ballistic transport, for all the typical
experimental parameters, when the voltage varies from $1$ to $5$ V and
sample lengths vary from $0.9$ to $5.9$ $\mathrm{\mu m}$ (see \cite%
{vandecasteele}), the corresponding ballistic times satisfy condition (\ref%
{g22}), $T_{bal}\ll \Delta t_{fin}$. This means that in realistic cases the
evolution of quantum states in course of the external field action satisfies
the restrictions which justify the approximation under consideration, so
that we can set $t_{fin}=t_{out}$.

Note that in the case when the field duration $t-t_{in}$ exceeds $\Delta
t_{fin}$ (that is unattainable for samples available at present), the mean
field $\bar{E}(t)$ becomes too weak in comparison with the external field $E$%
, and then the model with pair creation by a constant external field fails
to work. It is possible to consider the evolution of the mean field and the
vacuum mean current for time intervals greater than $\Delta t_{fin}$ if one
is able to calculate explicitly pair creation from vacuum due to a
time-dependent electric field, which could be a task for further study.
However, it is natural to assume that the residual effect of a weak electric
field cannot significantly change the asymptotic behavior achieved by the
time $t=t_{fin}$, which effectively means that the effective external field
duration is $T\approx t_{fin}$ $-t_{in}$. That is why in what follows we set
$t_{fin}=t_{out}$ for any $t_{out}$.

Substituting Eq.~(\ref{g19}) in Eq.~(\ref{g16}), we find the following
result for the density of the vacuum mean current:
\begin{equation}
\langle j^{1}(t)\rangle _{g}=\frac{e^{2}E}{2\pi \hbar \alpha }\left[
1-\epsilon(t)^{-2}\right] \, , \quad t\in (t_{in},t_{fin}] \, .  \label{g24}
\end{equation}
Thus, we have found self-consistent solutions for mean field and vacuum mean
current. From Eqs.~(\ref{g19}) and (\ref{g24}) we see that at $(t-t_{1}) \gg
\Delta t_{br}=\Delta t_{st}\pi /4\alpha $, the density of the vacuum mean
current and radiated electric field take asymptotic forms:
\begin{equation}
\langle j^{1}(t)\rangle _{g}\approx \frac{e^{2}E}{2\pi \hbar \alpha } \, ,
\quad E_{x}^{rad}\left( t\right) \approx -E \, .  \label{g25}
\end{equation}
Thus, the self-consistent system of the mean field and the vacuum mean
current at a given external electric field $E$ tends to a dynamic
equilibrium state in which the external field inside the graphene is
completely compensated by the radiated electric field. In this state the
particle production is stopped and the vacuum mean current saturates. Close
to this regime, the $I-V$ is almost linear. The momentum transferred by the
applied force is limited by the finite value $e\left\vert E\right\vert
\Delta t_{st}\pi /2\alpha$. Due to the discrete structure of the lattice,
formed by atoms at the carbon-carbon distance $a$, there is a natural
momentum cutoff of the order $\hbar /a$. For this reason, instead of being
restricted by the Bloch time $\Delta t_{B}$ mentioned in subsection \ref%
{SS4.1}, the applicability of the Dirac model is restricted, because of the
crystalline structure of graphene, by the field strength $E_{B}$,
\begin{equation*}
E_{B}\sim \left( \frac{4\alpha }{a}\right)^2 \frac{\hbar v_{F}}{e}\sim 3
\times 10^{7} \; \mathrm{V/m}.
\end{equation*}

Substituting Eq.~(\ref{g19}) into Eq.~(\ref{g28}), we can explicitly find
the diagonal elements $\langle T_{\mu \mu }\left( t\right) \rangle _{g}^{p}$%
. For example, the mean energy density reads:
\begin{equation}
\langle T_{00}\left( t\right) \rangle _{g}=8ev_{F}\rho \left\vert
E\right\vert ^{5/2}(\Delta t_{br})^2 \left[ \frac{1}{3}-\frac{1}{2\epsilon
(t) }+\frac{1}{6\epsilon(t)^3} \right] \, .  \label{g29}
\end{equation}
For $( t-t_{in}) \gg \Delta t_{br}$, we obtain from (\ref{g29}) the
following asymptotic form:
\begin{equation*}
\langle T_{00}\left( t\right) \rangle _{g}\approx \frac{8}{3}ev_{F}\rho
\left\vert E\right\vert ^{5/2}(\Delta t_{br})^2 \, .
\end{equation*}
One can see that $\langle T_{00}(t) \rangle_{g} \sim \left\vert E\right\vert
^{3/2}$ asymptotically for large times, while for small intervals $(
t-t_{in}) \ll \Delta t_{br}$ one has a dependence of the form $\langle
T_{00}(t) \rangle _{g}\sim \left\vert E\right\vert ^{5/2}$. The asymptotic
behavior of the element $\langle T_{22}(t) \rangle _{g}^{p}$ reads:
\begin{equation*}
\langle T_{22}(t) \rangle _{g}^{p}\approx \frac{\hbar \rho }{3\pi }
\left\vert E\right\vert ^{3/2} \, .
\end{equation*}
The behavior of the term $\langle T_{22}(t) \rangle _{g}^{c}$, given by (\ref%
{g28}), is completely determined by the field $\left\vert \bar{E}(t)
\right\vert $ at a given time instant, which means that this term tends to
zero asymptotically as $\epsilon(t)^{-3}$, and therefore can be neglected.
We see that $\langle T_{00}(t) \rangle  _{g}=\langle T_{11}(t) \rangle
_{g}\gg \langle T_{22}(t) \rangle _{g}$ during the whole evolution up to
asymptotically large values of the interval $( t-t_{in})$. Then, in this
approximation, we have that trace $\langle T_\mu^\mu(t) \rangle _{g}=0$, as it can be
expected for the equation of state for massless particles.

At $t=t_{out}$, the expressions (\ref{g24}) and (\ref{g29}) represent the
current density and energy density of created particles, respectively, and
the consistent mean electric field in graphene is given by Eq.~(\ref{g19}).
According to our problem setting, before the time moment $t_{out}$, we have
an unitary evolution of a pure state of the Dirac-Maxwell system. At the
initial time instant $t_{in}$, this pure state is the vacuum{\ for carriers}
and the coherent state of electromagnetic field with the initial mean value $%
E$. In the strong-field QED, \textquotedblleft the
measurement\textquotedblright\ which produces the decoherence (as a result
of which we obtain many-particle state of carriers and a final state of the
electromagnetic field specified by the mean value $\bar{E}\left(
t_{out}\right) $) occurs not necessary at the moment when the external field
switches off. This measurement can be done in any time instant after $t_{out}
$ because the further evolution of the system is trivial and mean values
remain unchanged. In the case of the finite graphene size and fixed constant
voltage this is not true because in such a case we adapt our model with the $%
T$-constant external field to the situation where the effective duration is
given, i.e., we set $T=T_{eff}$ and, therefore, by definition, we have $%
t_{out}=t_{in}+T$. (Note that one has to know the times $t_{in}$ and $T$
in a different experimental situation for the measurement of the time
evolution of the vacuum mean current, which has been been early proposed
\cite{allor} but not implemented yet.) In this case, the time instant $%
t_{out}$ of switching off the effective electric field is understood as the
effective time instant of the decoherence. We assume that the decoherence
stops evolution of pure states in a sufficiently short time, so that the
final mean electric current and EMT do not feel the effect of switching off
the effective electric field.

When the effective duration $T_{eff}$ is due to the time scale of
dissipation process $T_{dis}=l_{mfp}/v_{F}$, we set $%
t_{out}-t_{in}=T_{eff}=l_{mfp}/v_{F}$, while in the ballistic case $%
l_{mfp}=L_{x}$ and $T_{eff}=T_{bal}$. Then, using (\ref{g24}), we obtain the
$I-V$ curve in the following form:
\begin{align}
& \langle j^{1}(t_{out})\rangle _{g}=\frac{e^{2}V}{2\pi \hbar \alpha L_{x}}%
\left[ 1-\epsilon(t_{out})^{-2}\right] \, ,  \notag \\
& \epsilon( t_{out}) =1+\frac{2\alpha T_{eff}}{\pi \Delta t_{st}} \, , \quad
\frac{T_{eff}}{\Delta t_{st}}=\sqrt{\frac{e\left\vert V\right\vert
l_{mfp}^{2}}{\hbar v_{F}L_{x}}} \, .  \label{g31}
\end{align}

Example 1: Considering a sample with $L_{x}\sim 1\mu \mathrm{m}$\ and a
voltage $1\mathrm{V}$\ in the ballistic case, we have $T_{bal}=10^{-12}s$
and $\left\vert E\right\vert =10^{6}\mathrm{V/m}$. Alternatively, in the
case of dissipation, we can consider a sample of arbitrary length, where the
effective duration is $T_{eff}=10^{-12}s$, and the applied field strength is
$\left\vert E\right\vert =\left\vert V\right\vert /L_{x}=10^{6}\mathrm{V/m}$%
. In both cases, $2\alpha T_{eff}(\pi \Delta t_{st})^{-1}=0.18,$ and
condition (\ref{g8}) is satisfied. {But then} backreaction can be neglected
and the $I-V$\ curve from eq. (\ref{g31}) is close to $j\sim V^{3/2}$.

Example 2: Considering a sample with $L_{x}\sim 4\mu \mathrm{m}$\ and a
voltage $4\mathrm{V}$\ in the ballistic case, we have $2\alpha T_{eff}(\pi
\Delta t_{st})^{-1}=0.72$. Thus, condition (\ref{g8}) is not satisfied, and
then the backreaction contribution in eq. (\ref{g31}) is important, {%
resulting} in an almost linear $I-V$. {We see that our approach describe a
transition from a superlinear to a linear $I-V$ curve when passing from
low-mobility to high-mobility samples, in a manner similar to that observed
in \cite{vandecasteele}.}

Note that outside the graphene film there is an irradiated electromagnetic
plane wave $E_{x}^{rad}(t,z) =E_{x}^{rad}( t-\left\vert z\right\vert /c)$ of
linear polarization given explicitly by Eqs.~(\ref{g17}) and (\ref{g19}).
Therefore, the energy dissipation due to this irradiation must be taken into
account. In principle, such irradiation may be experimentally observed. It
is the radiation due to the time-dependence of the mean current. We have
calculated it nonperturbatively in the case when its electric component
strength is comparable to the external field strength. Thus, the mechanism
of the energy dissipation due to irradiation considered here differs
essentially from that discussed in Ref.~\cite{lewkowicz-11}. The authors of
the latter work have treated the radiation due to the electron-hole
recombination in the lowest order of the perturbation theory assuming large
enough densities of carriers. That is why the relevance of such a
consideration to the quantum transport close to the Dirac point is not clear.

\section{Summary\label{S5}}

In this summary, we briefly list the main new results obtained in the
article and add some relevant comments. These results are collected in three
blocks: I-General results in strong-field QED; II-Adaptation of the general
results to the Dirac model in the graphene; III-Analysis of some immediate
consequences to the graphene physics.

I. General results in strong-field QED

The one-loop renormalized mean current and EMT induced in the vacuum by the
$T$-constant external electric field are computed in the framework of
strong-field QED in spaces of arbitrary dimensions.

In the large $\tau$-limit, these quantities are represented as sums of local
contributions due to the vacuum polarization, ($\langle j^{\mu
}(t)\rangle _{\bot }^{c}$ and $\re \, \langle T_{\mu \nu
}(t)\rangle _{ren}^{c}$), and of global contributions due to the vacuum
instability, ($\re \, \langle j^{\mu }(t)\rangle ^{p}$ and $\re \, %
\langle T_{\mu \nu }(t)\rangle ^{p}$),
\begin{align*}
& \langle j^{\mu }(t)\rangle =\langle j^{\mu }(t)\rangle _{\bot }+\re \,%
\langle j^{\mu }(t)\rangle _{\Vert }^{p} \, , \\
& \langle T_{\mu \nu }(t)\rangle _{ren}=\re \, \langle T_{\mu
\nu }(t)\rangle _{ren}^{c}+\re \, \langle T_{\mu \nu }(t)\rangle ^{p} \, .
\end{align*}
These contributions are studied in detail. The vacuum polarization
contributions to the EMT are expressed via the real part of the one-loop
effective Euler-Heisenberg Lagrangian. In odd dimensions, unusual
peculiarities of the vacuum polarization emerge: along with the
longitudinal mean current, $\re \, \langle j^{\mu }(t)\rangle _{\Vert }^{p}$
(which behaves in the same way as in even dimensions), there appears a
transversal mean current, $\langle j^{\mu }(t)\rangle _{\bot }$. Its
components correspond to fermions of different chiralities (pseudo-spin)
moving in opposite directions. The sign of this current depends on the
fermion species. That is an indication that the Chern-Simons term is present
in a properly regularized effective action with odd number of fermion
species. In the case of an external electric field, this transverse mean
current depends essentially on the fermion mass (in contrast to the
corresponding contributions to probability amplitudes of processes); in
particular, it is zero in the massless case. Thus, the vacuum polarization
in the electric field in odd dimensions is qualitatively different for mean
values and for amplitudes of processes. This is the reason why the case of
electric-like field differs substantially from the case of magnetic-like
field. This fact is important in understanding properties of new materials
where the electronic structure is described by the Dirac model of $(2+1)$-d
fermions, in particular in the case of the interface transport in topological
insulators, where there is only one fermion species. Moreover, it should be
noted that even in the case of even number of fermion species with nonzero
masses, neutral electron-hole pairs induced due to the applied electric field
will be concentrated near lateral surfaces. Pairs with different chirality
are concentrated on opposite lateral surfaces (there is a polarization with
respect to the chirality). Otherwise, properties of these mean values in odd
and even dimensions do not differ essentially. In particular, the
longitudinal mean current components $\re \, \langle j^{\mu }(t)\rangle
_{\Vert }^{p}$ are linear functions of $T$, and the energy density $\langle
T_{00}(t)\rangle _{ren}$ and the pressure along the direction of the
electric field $\langle T_{11}(t)\rangle _{ren}$ are quadratic functions of
$T$.

II-Adaptation of the general results to the Dirac model in the graphene.

The general results described above are used in the non-perturbative
consideration of electronic and energy quantum transport in graphene at low
carrier density and low temperatures.

We have found: the time-dependence of the mean longitudinal current density
$\langle j^{1}(t)\rangle _{g}$, the mean EMT $\langle T_{\mu \mu }\left(
t\right) \rangle _{g}$, and the total number density of electron-hole pairs
$n_{g}^{cr}$ created by the electric field in the graphene. In these
calculations, we have used the strong-field approximation, the large
$\tau$-limit (when the duration $T$ of the external electric field is
sufficiently large), and the backreaction of created particles in screening
the external field was taken into account. All these quantities are obtained
for the case of an unitary evolution of a pure state of the Dirac-Maxwell
system. To adopt our consideration to the existing experimental situations,
where the length of graphene flakes is finite and a constant voltage is
applied, the time $T$ was replaced by some effective time $T_{eff}$, which
is determined by a certain decoherence. Assuming that this decoherence stops
the evolution of pure states in sufficiently short times one can neglect
effects of switching off the effective electric field on the final mean
electric current and EMT.

The new approach proposed for treating the quantum transport in graphene
allows to study unitary (coherent) evolution of system in the regime of
large duration of a strong external field, when the electrodynamic
backreaction is essential. In spite of the fact that we do not consider
statistical effects of the decoherence, we believe that exact analysis of
results of unitary evolution will be helpful as an initial stage for further
considering the thermalization of the system.

It should be noted that in some high-energy physics problems (QED and QCD)
in $(1+1)$- and $(3+1)$-dimensions the backreaction from the pair production
in electric fields was frequently discussed, in particular, in the framework
of WKB approximation (see e.g. \cite{NovStar80,KluMotEis98,Bloch+etal99})
and using the numerical approach (see e.g. \cite{mihaila}). Recently, the
exact solution for massless QED in $1+1$-dimensions was obtained, which by
the definition takes the back reaction into account completely \cite{ChuV10}%
. The authors of all these works considered only closed systems of fermions
interacting with a mean electromagnetic field, in which the total energy was
conserved. However, in considering the graphene in the framework of the
Dirac model the backreaction due to the pair creation from the vacuum has
never been studied. Our consideration of the backreaction in the graphene
due to the pair creation is completely new and, moreover, is essentially
different from that of the works cited above, because it \ concerns an open
system with a source that maintains a given voltage.

III-Analysis of some immediate consequences to the graphene physics.

In the large $\tau$-limit, the mean values $\langle j^{1}(t)\rangle _{g}$
and $\langle T_{\mu \mu }\left( t\right) \rangle _{g}$ are proportional to
the number density of pairs $r_{g}^{cr}$ of positive and negative charged
states excited by the electric field per unit time\textrm{. }At the end of
pure state evolution these quantities represent the current density and EMT
of created particles, respectively, so that $\langle j^{1}(t)\rangle
_{g}\sim \left\vert E\right\vert ^{3/2}$ and $\langle T_{00}\left( t\right)
\rangle _{g}=\langle T_{11}\left( t\right) \rangle _{g}\sim \left\vert
E\right\vert ^{5/2}$.

It is shown that there exists a parameter range $1\ll T/\Delta t_{st}\ll
\pi /4\alpha $ for which the external electric field of long duration
$T$ is a good approximation of the effective mean field inside the graphene.
Two corresponding characteristic time scales are established: one of them is
$\Delta t_{st}=\left( v_{F}e\left\vert E\right\vert /\hbar \right) ^{-1/2}$,
the time duration when the dc response goes from a linear to a non-linear
regime and the effect of the real electric field is indistinguishable
from the effect produced by a constant field; the other one is
$\Delta t_{br}=\Delta t_{st}\pi /4\alpha $, the time duration when
the backreaction becomes important. Such a specific regime is relevant to
describe the experimentally observed superlinearity $j\sim V^{3/2}$
of the $I-V$ in low mobility samples.

A generalization of the Dirac model with the $T$-constant electric field is
constructed that takes into account the backreaction of the mean current to
the applied electric field set by a constant voltage. It is shown that in
the case of graphene the electrodynamic backreaction is relevant and one can
see the interplay between two and three dimensions: the time-dependent mean
current is confined to the plane, but its radiation, forming the
backreaction on the plane, escapes to the three-dimensional space in the
form of linearly polarized plane electromagnetic waves. A self-consistent
solution of the Dirac-Maxwell set of equations for this generalized model is
found and the effective mean field and effective mean values of the current
and energy-momentum tensor are calculated. In this case the self-consistent
system of the mean field and the vacuum mean current tends to a dynamic
equilibrium state in which the external field inside the graphene is
completely compensated by the radiated electric field, the particle
production is stopped, and the vacuum mean current saturates. Close to this
regime, the $I-V$ is almost linear and the mean energy density reads:
$\langle T_{00}\left( t\right) \rangle _{g}\sim \left\vert E\right\vert
^{3/2} $. This mean field approximation is consistent if the inequality
$T/\Delta t_{st}\ll \pi ^{2}/8\alpha ^{2}$ holds true. This
restriction is much weaker than the one that determines consistency of the
external-field approximation.

We show that non-linear and linear $I-V$ experimentally observed in low and
high mobility samples, respectively, can be explained in the framework of
the presented consideration and that such a behavior is a consequence of the
fact that the conductivity in the graphene is essentially due to the pair
creation from vacuum by the applied electric field.

\begin{acknowledgments}
SPG acknowledges support of the program Bolsista CAPES/Brazil and thanks the
University of S\~{a}o Paulo for hospitality. DMG acknowledges the permanent
support of FAPESP and CNPq. SPG and DMG thank for partial support
by the Grant No. 14.B37.21.0911 of Russian Ministry of Science.
NY acknowledges support from CAPES (PRODOC
program).
\end{acknowledgments}

\end{document}